\documentclass[reprint,titlepage,amsmath,amssymb,aps,prd,nofootinbib,eqsecnum]{revtex4-2}

\usepackage{graphicx}
\usepackage[utf8]{inputenc}
\usepackage[T1]{fontenc}
\usepackage{newtxtext}
\usepackage{newtxmath}
\usepackage{microtype} 
\usepackage{physics} 
\usepackage{siunitx} 
\usepackage{tensor} 
\usepackage[svgnames]{xcolor} 
    \definecolor{accent}{HTML}{df2d16}
\usepackage[allcolors=accent,colorlinks,pdfusetitle,pdfauthor={Níckolas de Aguiar Alves and André G. S. Landulfo}]{hyperref} 
\usepackage{orcidlink} 

\newcommand{\hodge}{{\star}} 

\begin{document}
\title{Sound as a gauge theory and its infrared triangle}

\author{\firstname{Níckolas} de \surname{Aguiar Alves}\,\orcidlink{0000-0002-0309-735X}}
\email{alves.nickolas@ufabc.edu.br}

\author{André G. S. Landulfo\,\orcidlink{0000-0002-3717-4966}}
\email{andre.landulfo@ufabc.edu.br}

\affiliation{Center for Natural and Human Sciences, \href{https://ror.org/028kg9j04}{Federal University of ABC}, Avenida dos Estados 5001, Bangú, Santo André, São Paulo 09280-560, Brazil}

\begin{abstract}
Over the last few decades, a rich structure has been uncovered in the infrared sector of various field theories. This mostly comes through the connections between memory effects, asymptotic symmetries, and soft theorems (the ``infrared triangle''), which have been explored in much depth within high-energy physics. In this paper, we show how sound also admits an infrared triangle. We consider the linear perturbations of the Euler equations for a barotropic and irrotational fluid. We then show how low-frequency changes in an acoustic source can lead to lasting displacements of fluid particles. We proceed to write these linear perturbations in terms of a two-form potential---a Kalb--Ramond field, in the high-energy physics terminology. This phrases linear sound as a gauge theory. Standard techniques can then be used to probe the infrared structure of acoustics. We show how the memory effect relates to asymptotic symmetries in this dual formulation, and comment on how these notions can be connected to soft theorems. This exhibits an example of an infrared triangle in a condensed matter system and provides new pathways to the experimental detection of memory effects.
\end{abstract}

\maketitle

\section{Introduction}\label{sec: introduction}
    Humankind is interested in physics for many reasons, but it is easy to argue radiation is the most important among them. If we ever saw the stars at night, enjoyed the warmth of the Sun, listened to birds chirping in the woods, or even to binary black holes chirping in other galaxies, it was due to radiation emitted in its various forms.

    Astonishing as it is, we still have a lot to learn about radiation. Among the recent lessons, one of the most interesting is that radiative phenomena are not as ephemeral as they might have seem. As first noticed by \textcite{zeldovich1974RadiationGravitationalWaves}, the passage of a wave pulse can leave lasting effects. Their original setup concerned the emission of gravitational radiation by a cluster of stars, and they found the passage of a gravitational wave could leave a permanent displacement between two inertial probes located far away. This was eventually coined the ``memory effect'' and investigated by several authors \cite{braginsky1985KinematicResonanceMemory,braginsky1987GravitationalwaveBurstsMemory,christodoulou1991NonlinearNatureGravitation,blanchet1992HereditaryEffectsGravitational,thorne1992GravitationalwaveBurstsMemory,bieri2014PerturbativeGaugeInvariant,satishchandran2019AsymptoticBehaviorMassless,bieri2024GravitationalWaveDisplacement}. We now eagerly wait for gravitational wave observatories to test these predictions \cite{grant2023OutlookDetectingGravitationalwave,*grant2023ErratumOutlookDetecting,cogez2026DetectabilityGravitationalWaveMemory,zosso2026ClaimingDetectionGravitational}. In the meantime, new memories have been found in many other theories. For instance, electrodynamics now offers a new route for the experimental observation of memory effects \cite{bieri2013ElectromagneticAnalogGravitational,tolish2014ExaminationSimpleExample,tolish2014RetardedFieldsNull,bieri2024ExperimentMeasureElectromagnetic}.

    While the memory effect is interesting in and of itself, it has been linked to many other concepts that emerge at very large distances or very low energies. This is exemplified by the soft theorems in quantum field theories involving massless particles---see, for example, Refs. \cite{elvang2015ScatteringAmplitudesGauge,cheung2018TASILecturesScattering,strominger2018LecturesInfraredStructure,aguiaralves2026LecturesBondiMetzner}. These theorems express how a scattering amplitude behaves in the low-energy limit of one of its external legs. \textcite{strominger2016GravitationalMemoryBMS} showed how the memory effect in general relativity is related to a soft graviton theorem due to \textcite{weinberg1965InfraredPhotonsGravitons}. Similarly, the electromagnetic memory effect has been connected to a soft photon theorem \cite{pasterski2017AsymptoticSymmetriesElectromagnetic}. 

    In addition to this connection, there exists a third related topic in infrared physics, establishing an ``infrared triangle.'' The soft theorems mentioned above can be understood as dynamical consequences---more specifically, as Ward identities---of an underlying symmetry. Weinberg's soft graviton theorem, for example, has been shown \cite{strominger2014BMSInvarianceGravitational,campiglia2015AsymptoticSymmetriesGravity} to be an expression of invariance under Bondi--Metzner--Sachs transformations \cite{bondi1962GravitationalWavesGeneral,sachs1962AsymptoticSymmetriesGravitational,sachs1962GravitationalWavesGeneral}, which are symmetries of asymptotically flat spacetimes at infinity. See, for example, Ref. \cite{aguiaralves2026LecturesBondiMetzner} for an introduction. Similarly, the soft photon theorem was matched to new symmetries in quantum electrodynamics \cite{he2014NewSymmetriesMassless,campiglia2015AsymptoticSymmetriesQED}. Since these symmetries emerge when considering the long-distance behavior of these theories, they are known as asymptotic symmetries.

    The relations between memory, soft theorems, and asymptotic symmetries have led to many directions of investigation. These are often focused on the relations between symmetries and soft theorems, as these subjects are believed to be potential paths to a better understanding of quantum gravity \cite{pasterski2021CelestialHolography,pasterski2021LecturesCelestialAmplitudes,raclariu2021LecturesCelestialHolography,donnay2024CelestialHolographyAsymptotic,pasterski2025ChapterCelestialHolography}.
    
    In this paper, we choose to take a different perspective. Memory effects can be understood as a property of the wave equation in four-dimensions---something that can be seen by analyzing the derivation of memory in scalar field theories \cite{tolish2014RetardedFieldsNull,satishchandran2019AsymptoticBehaviorMassless}. While a massless Klein--Gordon field is usually understood in relativistic terms, it is mathematically described by the same sort of wave equation one could find in nonrelativistic physics. For example, linear perturbations in a fluid---popularly known as ``sound''---obey the very same equation. This is arguably the basis for analog gravity models \cite{unruh1981ExperimentalBlackHoleEvaporation,barcelo2011AnalogueGravity}. With this remark in mind, one may expect that sound obeys a memory effect as well. We show it does. Moreover, one may hope this memory effect is related to asymptotic symmetries and soft theorems, as in more fundamental theories such as general relativity and electrodynamics. We show it is. 

    Our approach is the following. Once the wave equation for acoustic perturbations has been derived---which is done in standard textbooks in acoustics and fluid mechanics \cite{kundu2025FluidMechanics,pierce2019AcousticsIntroductionIts,thorne2017ModernClassicalPhysics}---one can solve it with a source using a retarded Green's function to obtain the memory effect for a scalar field. The introduction of a source is standard when studying the generation of sonic perturbations \cite{kundu2025FluidMechanics,pierce2019AcousticsIntroductionIts,thorne2017ModernClassicalPhysics}, and the retarded Green's function approach is the standard method to derive the linear memory effect in general relativity \cite{braginsky1987GravitationalwaveBurstsMemory,bieri2024GravitationalWaveDisplacement,aguiaralves2026LecturesBondiMetzner}. In this way, a memory effect can be easily derived. 

    To understand the asymptotic symmetries of the system, we follow the results by \textcite{campiglia2018CanScalarsHave,campiglia2019ScalarAsymptoticCharges,francia2018TwoFormAsymptoticSymmetries,heissenberg2019TopicsAsymptoticSymmetries} concerning asymptotic symmetries for a relativistic scalar field. As done by them, we reformulate the scalar wave equation in terms of a two-form gauge field. The resulting field, known in high-energy physics as the Kalb--Ramond field \cite{kalb1974ClassicalDirectInterstring} or \(B\)-field \cite{polchinski1998IntroductionBosonicString,basile2025LecturesQuantumGravity,maccaferri2025IntroductionStringTheory}, is one of the simplest examples of \(p\)-form electrodynamics \cite{henneaux1986PFormElectrodynamics,lechner2018ClassicalElectrodynamicsModern}. A reformulation of hydrodynamics in terms of a two-form has been considered by other authors---see, for example, Refs. \cite{nambu1977StringsVorticesGauge,sugamoto1979DualTransformationAbelian,matsuo2021NoteDescriptionPerfect,fischer1999MotionQuantizedVortices,lund1976UnifiedApproachStrings} and references therein---although with different goals. For our purposes, the key remark is that \(p\)-form electrodynamics is a gauge theory, and thus its infrared structure can be understood in a now straightforward way. The infrared behavior of \(p\)-form fields has, in fact, been investigated by several authors \cite{divecchia2015SoftTheoremGraviton,donnay2023pFormsCelestialSphere,afshar2018AsymptoticSymmetriespForm,afshar2019StringMemoryEffect,esmaeili2020pFormGaugeFields,francia2024AsymptoticChargespforms,manzoni2026HigherorderpformAsymptotic,manzoni2026DualityAsymptoticCharges,romoli2024OrNTwoformAsymptotic}. This allows a prompt identification of asymptotic symmetries for sound, and then the connection to the previously discussed memory effect.

    At last, the infrared triangle is complete once a soft theorem is understood. This is most easily done in the scalar-field formulation. Because the sources in acoustics are akin to the method of images in electrostatics, the interpretation of a soft theorem is not as obvious as in high-energy theories such as general relativity and electrodynamics. We argue how the previous results by \textcite{campiglia2018CanScalarsHave,campiglia2019ScalarAsymptoticCharges,francia2018TwoFormAsymptoticSymmetries} can be used to understand acoustic perturbations in terms of soft theorems, but highlight how this should be interpreted differently from the high-energy counterparts. 

    Some authors have studied individual corners of a possible ``acoustic infrared triangle'' before. \textcite{datta2022InherentNonlinearityFluid} considered analogs of gravitational wave memory in Bose--Einstein condensates. They focus on flows that are approximately one-dimensional, and then study their nonlinear behavior. They find a permanent shift in the fluid density. These methods differ from ours, since our analysis considers flows in \(3+1\) dimensions, although we linearize the expressions. Furthermore, our memory effect is mainly codified in the permanent displacement of particles in the fluid. 
    
    Soft theorems in fluid systems have also been investigated. In fact, \textcite{cheung2023SoftPhononTheorems} studied soft phonon theorems for a variety of condensed matter systems. We are not aware, though, of previous investigations of asymptotic symmetries for sound, much less of a pursuit for an acoustic infrared triangle. We note, however, that \textcite{perez2024FractonInfraredTriangle} discussed an infrared triangle for fractons in condensed matter physics. In this sense, our results provide one of the first examples of an infrared triangle in a condensed matter system. 

    The paper is organized in the following way. Section \ref{sec: linear-acoustics} reviews the basic concepts of linear sound we will need in the remaining sections. We first derive the wave equation for acoustic perturbations in the simple case of a homogeneous and quiescent medium. We discuss how sound waves can be generated by time-dependent boundary-conditions, and how these boundary conditions can be traded by localized sources using the method of images commonly employed in electrostatics. Using the retarded Green's function for the wave operator, we derive a memory effect predicting the lasting displacement of fluid particles due to changes on the sources.

    To proceed with the infrared triangle, it is more convenient to rephrase linear sound in terms of the Kalb--Ramond theory. This is done in Sec. \ref{sec: sound-as-gauge}. This rephrasing can be seen as a formulation of analog gravity models in terms of a two-form. We thus perform the construction using essentially the same hypotheses as in the seminal paper by \textcite{unruh1981ExperimentalBlackHoleEvaporation}. Namely, the construction considers an inviscid, irrotational, and barotropic fluid. Nevertheless, it is not necessary to make assumptions about homogeneity or quiescence. We also discuss a few general aspects of this reformulation. 

    Before proceeding to the discussion of the infrared triangle for acoustics, we review the case of Maxwell electrodynamics in Sec. \ref{sec: electrodynamics}. This allows a clear understanding of what our goals in the gauge theory for sound will be, and clarifies the difficulties and strategies to be considered in the following sections. Sections \ref{sec: linear-acoustics}, \ref{sec: sound-as-gauge}, and \ref{sec: electrodynamics} are mostly independent from each other, but their discussions converge on Sec. \ref{sec: memory-kalb-ramond}. There we rederive the memory effect in acoustics in the two-form formulation by following the same approach used in electrodynamics. The effect is then connected to the asymptotic symmetries for the two-form field in Sec. \ref{sec: asymptotic-symmetries}. The connection of these asymptotic symmetries to soft theorems had already been clarified in the work of \textcite{campiglia2018CanScalarsHave,campiglia2019ScalarAsymptoticCharges,francia2018TwoFormAsymptoticSymmetries}. This connection, and the relation between soft theorems and the acoustic memory effect, are briefly considered in Sec. \ref{sec: soft-theorem}. The results of the paper are recollected and further discussed in Sec. \ref{sec: discussion}, where we also comment on future prospects. Appendix \ref{app: differential-forms} briefly reviews some concepts about differential forms and de Rham cohomology that are useful throughout the paper. 
    
    Expressions involving differential geometry often employ abstract index notation \cite{wald1984GeneralRelativity}, and we follow the \({-}{+}{+}{+}\) metric signature convention. Lowercase Latin indices from the beginning of the alphabet (\(a\), \(b\), \(c\), \ldots) denote abstract indices in spacetime, while Greek indices denote coordinate indices. Lowercase Latin indices \(i\), \(j\), \(k\), \ldots are occasionally used to denote indices on a three-dimensional manifold, and uppercase Latin indices \(A\), \(B\), \(C\), \ldots are used to denote indices on the two-sphere. No distinction between abstract and coordinate indices is made in either of these cases. The round metric on the unit two-sphere is given by \(\tensor{\gamma}{_A_B}\), with Levi-Civita connection \(\tensor{\mathcal{D}}{_A}\) and Laplacian \(\mathcal{D}^2 = \tensor{\mathcal{D}}{^A}\tensor{\mathcal{D}}{_A}\). Section \ref{sec: electrodynamics}, on electrodynamics, is written in Heaviside--Lorentz units with the speed of light set to \(c=1\). Sections \ref{sec: memory-kalb-ramond}, \ref{sec: asymptotic-symmetries}, and \ref{sec: soft-theorem} work in units where the fluid background density is \(\rho_0 = 1\) and the speed of sound is \(c = 1\).

\section{Linear Sound and Acoustic Memory}\label{sec: linear-acoustics}
    We begin by considering the linearized theory of sound. We will first derive the wave equation for acoustic perturbations from the basic equations describing ideal fluids in the simplest case of a homogeneous and quiescent background. Then we will argue how one can use the method of images to introduce sources. This discussion is mostly based on Refs. \cite{kundu2025FluidMechanics,pierce2019AcousticsIntroductionIts,thorne2017ModernClassicalPhysics}. We will end the section by deriving an acoustic memory effect using the retarded Green's function for the wave equation.

    \begin{figure*}[tbph]
        \centering
        \null\hfill\includegraphics{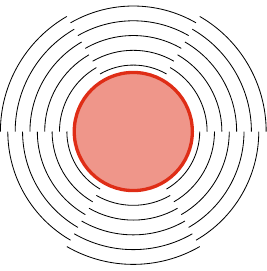}\hfill\includegraphics{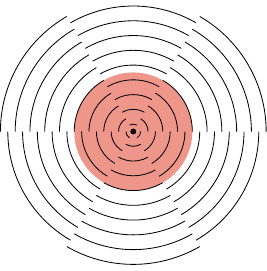}\hfill\null
        \caption{Left: a moving surface, such as a sphere with oscillating radius, disturbs the fluid around it and generates sound waves. Right: using the method of images, we can understand the sound waves as being generated by ``image sources'' inside the surface. We solve the problem in a larger region, with the understanding that the interior of the original surface (shaded region) is to be considered unphysical.}
        \label{fig: image-sources}
    \end{figure*}

    \subsection{Linear acoustic perturbations and sources}
        The equations ruling the behavior of an ideal fluid are the mass conservation and Euler equations
        \begin{subequations}
            \begin{gather}
                \pdv{\rho}{t} + \div(\rho \vb{v}) = 0, \\
                \rho\qty[\pdv{\vb{v}}{t} + (\vb{v} \vdot \vb*{\nabla})\vb{v}] = - \grad P - \rho \grad\Phi,
            \end{gather}
        \end{subequations}
        where \(\rho\) is the fluid's mass density, \(\vb{v}\) its velocity, \(P\) its pressure, and \(\Phi\) is a potential for an external force. For example, \(\Phi\) could be the Newtonian gravitational potential. 
        
        Our goal is to study linearized perturbations of the fluid flow. Nevertheless, we are not interested only in the free flow, but also in the generation of sound waves. While sound can be produced through various mechanisms, a simple one is to introduce time-dependent boundary conditions when solving the equations describing the fluid flow. For instance, if the skin of a drum is vibrating in a certain known way, this introduces boundary conditions on the airflow close to the drum. The time-dependent boundary conditions create disturbances on the fluid, which propagate as sound waves. 
    
        With this in mind, let us consider the example of a pulsating sphere---discussed, for example, in Refs. \cite{pierce2019AcousticsIntroductionIts,thorne2017ModernClassicalPhysics}. We assume a spherically symmetric fluid flow, with a sphere with oscillating radius put at the origin. This sphere could serve as a simplified model of, for example, the croaking of a frog, or the ringing of a bell. We will assume the time-dependence of the sphere's radius to be known in advance. Namely, we assume the sphere to have a radius \(a(t)\). The fluid is not allowed to cross the surface of the sphere, and thus the velocity of the surface determines the velocity of the fluid. Mathematically, we impose 
        \begin{equation}\label{eq: boundary-condition-sphere}
            \vb{v}|_{r=a(t)} = \dot{a}(t) \vu{r},
        \end{equation}
        where we recognize \(\dot{a}(t)\) as the radial velocity of the surface of the sphere. 
        
        Notice Eq. \eqref{eq: boundary-condition-sphere} imposes a boundary condition on a moving surface. This makes the problem particularly difficult to analyze. We can, however, try following a different route. Notice that the boundary condition implies, through the divergence theorem,
        \begin{subequations}\label{eq: volume-integral-velocity}
            \begin{align}
                \int_{r \leq a(t)} \div\vb{v} \dd{V} &= \oint_{r = a(t)} (\vb{v} \vdot \vu{r}) r^2 \dd{\Omega}, \\
                &= \oint \dot{a}(t) a(t)^2 \dd{\Omega}, \\
                &= 4 \pi a(t)^2 \dot{a}(t), \\
                &= \dv{t}\qty(\frac{4\pi a(t)^3}{3}).
            \end{align}
        \end{subequations}
        We can interpret this result in terms of ``volume sources'' inside the sphere. The continuity equation forbids these sorts of sources, but notice the interior of the sphere is not a physical region. We are not trying to model the fluid flow in there. Therefore, we can understand these sources as ``image sources,'' in complete analogy with the method of images used in electrostatics. Instead of handling the difficult boundary conditions on a moving surface, we introduce unphysical ``image sources'' inside the sphere as a trick to simplify our calculations and reproduce the solution outside the sphere. This is illustrated in Fig. \ref{fig: image-sources}. For more details, see, for example, Refs. \cite{kundu2025FluidMechanics,pierce2019AcousticsIntroductionIts}. This method was developed in more generality by \textcite{ffowcswilliams1969SoundGenerationTurbulence}, who consider nonlinear and viscous effects as well. See also Refs. \cite{farassat1975TheoryNoiseGeneration,glegg2017AeroacousticsLowMach,goldstein1976Aeroacoustics,howe2014AcousticsAerodynamicSound} for reviews.
        
        We thus proceed by manually adding a source term to the equations of motion of the fluid, 
        \begin{subequations}
            \begin{gather}
                \pdv{\rho}{t} + \div(\rho \vb{v}) = \rho q, \\
                \rho\qty[\pdv{\vb{v}}{t} + (\vb{v} \vdot \vb*{\nabla})\vb{v}] = - \grad P - \rho \grad\Phi.
            \end{gather}
        \end{subequations}
        Above, \(q\) is a function of time and space modeling (some of the) sources of disturbances in the fluid. We can approximate the case of an pulsating sphere by picking, for example, 
        \begin{equation}
            q(t, \vb{r}) = 4 \pi a(t)^2 \dot{a}(t) \delta^{(3)}(\vb{r}).
        \end{equation}
        This choice reproduces the integral in Eq. \eqref{eq: volume-integral-velocity}. A more detailed analysis could further develop the choice of source, but this model can already yield good approximations \cite{pierce2019AcousticsIntroductionIts,frost1975AcousticRadiationSurfaces,strasberg1956GasBubblesSources}.
    
        Since our interest is in describing linearized sound waves, the next step is to consider perturbation theory. For this purpose, we will understand the source terms as generating the perturbations, but not affecting the underlying medium. For the sake of simplicity, we will also assume the background medium to be homogeneous and quiescent, meaning all background quantities are time and space-independent and the equilibrium velocity is zero. To do so, we choose \(\Phi = 0\). Hence, for any fluid variable \(\psi\), we will write
        \begin{equation}
            \psi(t,\vb{r}) = \psi_0 + \var\psi(t,\vb{r}),
        \end{equation}
        where \(\psi_0\) is the (constant) equilibrium solution, while \(\var\psi(t,\vb{r})\) is the perturbation. Since \(\vb{v}_0 = \vb{0}\), we may write \(\vb{v} = \var{\vb{v}}\) without the risk of confusion. The linearized equations of motion are then
        \begin{subequations}
            \begin{gather}
                \pdv{\var{\rho}}{t} + \rho_0 \div\vb{v} = \rho_0 q, \\
                \rho_0\pdv{\vb{v}}{t} + \grad \var{P} = \vb{0}.
            \end{gather}
        \end{subequations}
        We have two differential equations, but three unknown variables---the latter being the perturbations on velocity, pressure, and density. The missing equation is an equation of state. Hence, we will assume the perturbations to be adiabatic,
        \begin{equation}\label{eq: adiabatic-perturbation}
            \var{P} = \qty(\pdv{P}{\rho})_s \var{\rho},
        \end{equation}
        where the derivative is evaluated at the background solution. The adiabatic speed of sound is then defined as 
        \begin{equation}
            c^2 = \qty(\pdv{P}{\rho})_s.
        \end{equation}
        The equations of motion for the perturbations can thus be written as 
        \begin{subequations}
            \begin{gather}
                \frac{1}{c^2}\pdv{\var{P}}{t} + \rho_0 \div\vb{v} = \rho_0 q, \\
                \rho_0\pdv{\vb{v}}{t} + \grad \var{P} = \vb{0}. \label{eq: linearized-eom-vorticity-conservation}
            \end{gather}
        \end{subequations}
        
        Equation \eqref{eq: linearized-eom-vorticity-conservation} implies 
        \begin{equation}
            \pdv{t}\qty(\curl{\vb{v}}) = \vb{0}.
        \end{equation}
        In other words, the vorticity \(\curl{\vb{v}}\) of the fluid flow is conserved. We will assume it to be initially zero, which automatically implies it is always zero. Therefore, the Helmholtz theorem implies we can obtain a potential function \(\phi\) such that
        \begin{equation}\label{eq: velocity-from-potential}
            \vb{v} = \grad \phi.
        \end{equation}
        Up to a redefinition of this potential, we see the linearized equations of motion now read 
        \begin{subequations}
            \begin{gather}
                \frac{1}{c^2}\pdv{\var{P}}{t} + \rho_0 \laplacian\phi = \rho_0 q, \label{eq: pde-pressure-phi-source} \\
                \rho_0\pdv{\phi}{t} + \var{P} = 0. \label{eq: pressure-from-potential}
            \end{gather}
        \end{subequations}
        
        We can differentiate Eq. \eqref{eq: pressure-from-potential} with respect to time and replace it in Eq. \eqref{eq: pde-pressure-phi-source}. We then find that 
        \begin{equation}\label{eq: wave-equation-phi}
            -\frac{1}{c^2}\pdv[2]{\phi}{t} + \laplacian\phi = q.
        \end{equation}
        We thus have found a wave equation for the potential \(\phi\). Once we solve it, we can obtain all other fluid variables through Eqs. \eqref{eq: adiabatic-perturbation}, \eqref{eq: velocity-from-potential}, and \eqref{eq: pressure-from-potential}. We recall that the source term \(q\) was introduced to avoid dealing with complicated boundary conditions. Thus, when solving the wave equation, we will search for a (possibly distributional) solution that holds across all space. This solution may not be physical in some regions of space, precisely due to the boundary conditions we are avoiding for simplicity. In the example of a pulsating sphere, the solution for \(\phi\) is unphysical inside the sphere.
    
    \subsection{Acoustic memory}
        We are interested only in the sound waves generated by the source term \(q\). Therefore, we can solve Eq. \eqref{eq: wave-equation-phi} promptly with the aid of the retarded Green's function for the wave equation,
        \begin{equation}\label{eq: retarded-solution-phi}
            \phi(t,\vb{r}) = - \frac{1}{4\pi} \int \frac{q\qty(t_{\text{ret}}, \vb{r}')}{\norm{\vb{r} - \vb{r}'}} \dd[3]{r'}.
        \end{equation}
        Above, the retarded time \(t_{\text{ret}}\) is given by 
        \begin{equation}
            t_{\text{ret}} = t - \frac{\norm{\vb{r} - \vb{r}'}}{c}.
        \end{equation}
    
        Our original setup considered boundary conditions near the origin, and the image source \(q\) is meant to model these boundary conditions. Therefore, for any instant \(t\), we shall assume \(q\) to have compact support in a vicinity of the origin and our observation point \(\vb{r}\) to be very far away. We may then expand \(\norm{\vb{r} - \vb{r}'}\) according to
        \begin{subequations}
            \begin{align}
                \norm{\vb{r} - \vb{r}'} &= r \norm{\vu{r} - \frac{\vb{r}'}{r}}, \\
                &= r \qty[1 - \frac{\vb{r}' \vdot \vu{r}}{r} + \order{\frac{1}{r^2}}], \\
                &= r - \vb{r}' \vdot \vu{r} + \order{\frac{1}{r}}.
            \end{align}
        \end{subequations}
        By using the above expansion in Eq. \eqref{eq: retarded-solution-phi} we find
        \begin{equation}\label{eq: large-r-integral-phi}
            \phi(t,\vb{r}) = - \frac{1}{4\pi r} \int q\qty(t-\frac{r}{c}+\frac{\vb{r}' \vdot \vu{r}}{c}, \vb{r}') \dd[3]{r'} + \order{\frac{1}{r^2}},
        \end{equation}
        where it is implicitly assumed that \(\norm{\vb{r}'} \ll r\) for all points \(\vb{r}'\) in the support of \(q(t,\vb{r}')\). This is completely analogous to radiation-zone calculations in electrodynamics---see, for example, Ref. \cite{wald2022AdvancedClassicalElectromagnetism}. We can summarize this result as 
        \begin{equation}\label{eq: general-solution-phi}
            \phi(t,\vb{r}) = - \frac{Q\qty(t-r/c,\vu{r})}{4\pi r} + \order{\frac{1}{r^2}},
        \end{equation}
        for some function \(Q\). Expanding the angular dependence of \(Q\) in spherical harmonics would lead us to a multipole expansion. For example, the monopole is the spherically symmetric term. Notice that, in acoustics, there are no symmetries enforcing the conservation of the acoustic monopole.
        
        Once we know \(\phi\), we can obtain other quantities. For example, Eqs. \eqref{eq: velocity-from-potential} and \eqref{eq: pressure-from-potential} allow us to extract the velocity and pressure from Eq. \eqref{eq: general-solution-phi} as
        \begin{subequations}
            \begin{gather}
                \vb{v}(t,\vb{r}) = \frac{\dot{Q}\qty(t - r/c, \vu{r})}{4 \pi c r}\vu{r} + \order{\frac{1}{r^2}} \\
                \intertext{and}
                P(t,\vb{r}) = P_0 + \frac{\rho_0 \dot{Q}\qty(t - r/c, \vu{r})}{4 \pi r} + \order{\frac{1}{r^2}},
            \end{gather}
        \end{subequations}
        respectively. 
    
        Let us consider how the position of a fluid particle changes before and after the passage of a sound wave. This is quantified by the Lagrangian displacement \(\vb*{\xi}\). More specifically, if the unperturbed position of a particle is \(\vb{r}\), then the perturbation takes it to \(\vb{r} + \vb*{\xi}(t,\vb{r})\). Notice 
        \begin{equation}
            \pdv{\vb*{\xi}}{t} = \vb{v}
        \end{equation}
        at linear order in the perturbations. Between early and late times, the change in the Lagrangian displacement is 
        \begin{subequations}\label{eq: Delta-xi-from-phi}
            \begin{align}
                \Delta \vb*{\xi}(\vb{r}) &= \frac{\qty[\left.Q(\vu{r})\right|_{+\infty} - \left.Q(\vu{r})\right|_{-\infty}]}{4\pi c r}\vu{r} + \order{\frac{1}{r^2}}, \\
                &= - \frac{\Delta\phi\,\vu{r}}{c} + \order{\frac{1}{r^2}}.
            \end{align}
        \end{subequations}
    
        For \(\Delta \vb*{\xi}(\vb{r})\) to be well-defined, it is necessary that \(\dot{Q}\) vanishes at early and late times. If this does not happen, but \(\ddot{Q}\) vanishes, then we can still see that 
        \begin{subequations}
            \begin{gather}
                \Delta\vb{v}(\vb{r}) = \frac{1}{4\pi c r}\qty[\eval{\dv{Q}{t}}_{+\infty} - \eval{\dv{Q}{t}}_{-\infty}]\vu{r} + \order{\frac{1}{r^2}} \\
                \intertext{and}
                \Delta P(\vb{r}) = \frac{\rho_0}{4 \pi r} \qty[\eval{\dv{Q}{t}}_{+\infty} - \eval{\dv{Q}{t}}_{-\infty}] + \order{\frac{1}{r^2}}.
            \end{gather}
        \end{subequations}
    
        In other words, depending on whether \(\dot{Q}\) or \(\ddot{Q}\) vanish at early and late times, the Lagrangian displacement, velocity, or pressure present an infrared memory effect, much like the ones predicted in gravity and gauge theories \cite{aguiaralves2026LecturesBondiMetzner,bieri2013ElectromagneticAnalogGravitational,bieri2024ExperimentMeasureElectromagnetic,bieri2024GravitationalWaveDisplacement,blanchet1992HereditaryEffectsGravitational,braginsky1985KinematicResonanceMemory,braginsky1987GravitationalwaveBurstsMemory,christodoulou1991NonlinearNatureGravitation,satishchandran2019AsymptoticBehaviorMassless,thorne1992GravitationalwaveBurstsMemory,tolish2014ExaminationSimpleExample,tolish2014RetardedFieldsNull,zeldovich1974RadiationGravitationalWaves,bieri2014PerturbativeGaugeInvariant,grant2023OutlookDetectingGravitationalwave}. In particular, general relativity also allows a ``velocity-coded memory effect'' \cite{grishchuk1989GravitationalWavePulses,bieri2024GravitationalWaveDisplacement}. Due to the conservation laws of general relativity, gravitational wave memory is associated to changes in the second time-derivative of the mass quadrupole---see, e.g., Refs. \cite{aguiaralves2026LecturesBondiMetzner,bieri2024GravitationalWaveDisplacement}. Dipole variations would violate energy-momentum conservation. Meanwhile, electromagnetic memory is associated to changes in the time-derivative of the electric dipole. Monopole variations would violate charge conservation. Since there are no conservation laws associated with acoustic sources, it follows that monopole variations are sufficient to source memory. 
        
        In the example of a pulsating sphere, \(Q(t)\) is the time-derivative of the sphere's volume. Hence, a finite \(Q\) at late times would mean the sphere is eternally expanding at a constant rate of change of volume. In practical experiments, however, it is not necessary to consider the expressions at infinite times---as discussed in Ref. \cite{bieri2024ExperimentMeasureElectromagnetic}, it should also be possible to observe memory effects at finite times. In this acoustic scenario, it suffices that the sphere continues expanding for a sufficiently long, but finite, time. Here, ``sufficiently long'' is meant relative to the intermediate, oscillating phase.

        One could now wonder how expressive the acoustic displacement memory is. Let us consider the case of a spherical source. If the sphere starts at fixed radius, but ends in an expansion with constant acoustic monopole, then Eq. \eqref{eq: Delta-xi-from-phi} tells us that the magnitude of the displacement will be 
        \begin{equation}
            \Delta \xi \approx \frac{Q|_{+\infty}}{4 \pi c r}.
        \end{equation}
        For a sphere of radius \(a(t)\), this means 
        \begin{equation}
            \Delta \xi \approx \frac{a(+\infty)^2 \dot{a}(+\infty)}{c r},
        \end{equation}
        where we assume \(a(t)\) to be roughly constant at late times, and similarly for \(\dot{a}(t)\). This assumption will be reasonable for sufficiently fast measurements, where ``sufficiently fast'' depends on the value of \(\dot{a}\).
    
\section{Sound as a Gauge Theory}\label{sec: sound-as-gauge}
    Much of the recent theoretical work in memory effects has focused on the connections to asymptotic symmetries and soft theorems in quantum field theory. See, for example, Refs. \cite{aguiaralves2026LecturesBondiMetzner,strominger2018LecturesInfraredStructure} for pedagogical introductions. Given the acoustic memory effect derived above, it is natural to search for asymptotic symmetries in the theory of linearized sound. To do so, it is convenient to reformulate the acoustic perturbations in terms of a two-form field. We will thus rely considerably on the theory of differential forms, which is discussed in Refs. \cite{frankel2012GeometryPhysicsIntroduction,gorodski2020SmoothManifolds,lechner2018ClassicalElectrodynamicsModern,lee2012IntroductionSmoothManifolds,nakahara2003GeometryTopologyPhysics,tu2011IntroductionManifolds}, for example. In App. \ref{app: differential-forms} we provide a brief summary of the main results we will need.

    So far, we described sound in terms of a scalar field, meaning we are now interested in studying the asymptotic symmetries of a scalar field. This was considered by \textcite{campiglia2018CanScalarsHave,campiglia2019ScalarAsymptoticCharges,francia2018TwoFormAsymptoticSymmetries,heissenberg2019TopicsAsymptoticSymmetries}---see also Ref. \cite{aneesh2022CelestialHolographyLectures} for a review. In Refs. \cite{campiglia2019ScalarAsymptoticCharges,francia2018TwoFormAsymptoticSymmetries,heissenberg2019TopicsAsymptoticSymmetries}, it is argued that asymptotic symmetries for scalar fields in four dimensions can be understood by reformulating the theory in terms of a two-form. While we will discuss asymptotic symmetries later on Sec. \ref{sec: asymptotic-symmetries}, let us now focus on how a two-form can appear when describing acoustics. 
    
     Let us consider the linearization of the system of equations
    \begin{subequations}
        \begin{gather}
            \pdv{\rho}{t} + \div(\rho \vb{v}) = 0, \\
            \rho\qty[\pdv{\vb{v}}{t} + (\vb{v} \vdot \vb*{\nabla})\vb{v}] = - \grad P - \rho \grad\Phi.
        \end{gather}
    \end{subequations}
    We further assume that the fluid is irrotational (\(\curl{\vb{v}} = \vb{0}\)) and barotropic (\(P = P(\rho)\)). In particular, this implies the perturbations are adiabatic, and we denote the adiabatic speed of sound by \(c\) as before. Notice \(c\) may depend on time and space, for we make no further assumptions on homogeneity or quiescence at this stage. Upon linearization, the equations of motion become
    \begin{subequations}
        \begin{gather}
            \pdv{\rho_0}{t} + \div(\rho_0 \vb{v}_0) = 0, \\
            \rho_0\qty[\pdv{\vb{v}_0}{t} + (\vb{v}_0 \vdot \vb*{\nabla})\vb{v}_0] = - \grad P_0 - \rho_0 \grad\Phi,
        \end{gather}
    \end{subequations}
    for the unperturbed quantities, and 
    \begin{subequations}
        \begin{gather}
            \pdv{\var{\rho}}{t} + \div(\var{\rho} \vb{v}_0 + \rho_0 \var{\vb{v}}) = 0, \label{eq: perturbed-mass-conservation} \\
            \rho_0\qty[\pdv{\var{\vb{v}}}{t} + (\var{\vb{v}} \vdot \vb*{\nabla})\vb{v}_0 + (\vb{v}_0 \vdot \vb*{\nabla})\var{\vb{v}}] = \frac{\var{\rho} \grad P_0 - \rho_0 \grad \var{P}}{\rho_0} \label{eq: perturbed-euler}
        \end{gather}
    \end{subequations}
    for the perturbed quantities. It is known that the linear acoustic perturbations we are considering can be described in terms of a massless scalar field in a Lorentzian spacetime \cite{unruh1981ExperimentalBlackHoleEvaporation,barcelo2011AnalogueGravity}. While this is not the route we will follow here, it will be convenient to notice this description employs the ``acoustic metric''
    \begin{equation}\label{eq: acoustic-metric}
        \dd{s}^2 = \frac{\rho_0}{c}[-c^2 \dd{t}^2 + (\dd{\vb{x} - \vb{v}_0 \dd{t}})^2].
    \end{equation}
    Notice the speed of sound \(c\) is a function in this expression, not necessarily a constant.

    Let us define the three-form \(\mathcal{H}\) (the ``acoustic field strength'') by 
    \begin{multline}\label{eq: H-form}
        \mathcal{H} = \var{\rho} \dd{x} \wedge \dd{y} \wedge \dd{z} \\ - \frac{1}{2} \epsilon_{ijk} (\var{\rho} v_0^i + \rho_0 \var{v}^i) \dd{t} \wedge \dd{x^j} \wedge \dd{x^k},
    \end{multline}
    where \(\epsilon_{ijk}\) is the three-dimensional volume form in Euclidean space. Notice then that\footnote{The Supplemental Material accompanying this preprint includes a \textsc{Mathematica} \cite{wolframresearch2026Mathematica150} notebook used to perform the manipulations with the acoustic field strength \(\mathcal{H}\). The code is written with the \textsc{OGRe} package \cite{shoshany2021OGReObjectOrientedGeneral}.}
    \begin{equation}
        \dd{\mathcal{H}} = \qty(\pdv{\var{\rho}}{t} + \div(\var{\rho} \vb{v}_0 + \rho_0 \var{\vb{v}})) \dd{t} \wedge \dd{x} \wedge \dd{y} \wedge \dd{z}.
    \end{equation}
    Hence, \(\mathcal{H}\) is a closed form if, and only if, Eq. \eqref{eq: perturbed-mass-conservation} holds. We will now show that Eq. \eqref{eq: perturbed-euler} can be expressed similarly.

    In the acoustic spacetime [with metric given by Eq. \eqref{eq: acoustic-metric}], we can write the Hodge dual  of \(\mathcal{H}\) (see App. \ref{app: differential-forms}) as 
    \begin{equation}
        \hodge \mathcal{H} = - \qty(\vb{v}_0 \vdot \var{\vb{v}} + \frac{c^2 \var{\rho}}{\rho_0})\dd{t} + \var{\vb{v}} \vdot \dd{\vb{x}}.
    \end{equation}
    Because the perturbations are adiabatic, we can also write
    \begin{equation}
        \hodge \mathcal{H} = - \qty(\vb{v}_0 \vdot \var{\vb{v}} + \frac{\var{P}}{\rho_0})\dd{t} + \var{\vb{v}} \vdot \dd{\vb{x}}.
    \end{equation}
    Due to our assumption that \(\curl{\vb{v}} = \vb{0}\), it follows that \(\curl{\var{\vb{v}}} = \vb{0}\). We may thus compute the exterior derivative of \(\hodge \mathcal{H}\) and find it to be 
    \begin{equation}
        \dd\hodge \mathcal{H} = - \qty[\pdv{\var{\vb{v}}}{t} + \grad\qty(\vb{v}_0 \vdot \var{\vb{v}} + \frac{\var{P}}{\rho_0})]\vdot\dd{\vb{x}} \wedge \dd{t}.
    \end{equation}
    Since we assume the flow to be irrotational, it follows from standard vector calculus identities that
    \begin{multline}\label{intermediate}
        \pdv{\var{\vb{v}}}{t} + \grad\qty(\vb{v}_0 \vdot \var{\vb{v}} + \frac{\var{P}}{\rho_0}) \\ = \pdv{\var{\vb{v}}}{t} + \qty(\vb{v}_0 \vdot \grad)\var{\vb{v}} + \qty(\var{\vb{v}}\vdot\grad)\vb{v}_0 + \frac{\grad\var{P}}{\rho_0} - \frac{\var{P} \grad\rho_0}{\rho_0^2}.
    \end{multline}
    By using Eq. \eqref{eq: perturbed-euler} we can rewrite the right-hand side of Eq. \eqref{intermediate} as
    \begin{equation}
        \pdv{\var{\vb{v}}}{t} + \grad\qty(\vb{v}_0 \vdot \var{\vb{v}} + \frac{\var{P}}{\rho_0}) = \frac{\var{\rho} \grad P_0 - \var{P} \grad\rho_0}{\rho_0^2}.
    \end{equation}
    The right-hand side vanishes due to the assumption of a barotropic equation of state, which enforces both \(\grad P_0 = c^2 \grad \rho_0\) and \(\var{P} = c^2 \var{\rho}\). Hence, we find
    \begin{equation}
        \dd\hodge \mathcal{H} = 0.
    \end{equation}

    We have thus learned that the linear perturbations of an inviscid, irrotational, barotropic fluid flow are described by the equations of motion
    \begin{equation}\label{eq: form-field-equations}
        \dd \mathcal{H} = 0 \qq{and} \dd\hodge \mathcal{H} = 0,
    \end{equation}
    where we recall that the ``field strength'' three-form \(\mathcal{H}\) is given by Eq. \eqref{eq: H-form} and the Hodge dual is meant with respect to the acoustic metric \eqref{eq: acoustic-metric}. This is a dual formulation of the fluid-gravity analogy first proposed by \textcite{unruh1981ExperimentalBlackHoleEvaporation}, which eventually developed into the field of analog gravity \cite{barcelo2011AnalogueGravity}.  A very interesting property of Eq. \eqref{eq: form-field-equations} is the resemblance to Maxwell's equations in the absence of sources. In fact, the equations we found are precisely the equations of motion for two-form electrodynamics \cite{kalb1974ClassicalDirectInterstring,henneaux1986PFormElectrodynamics} in the absence of sources---see, e.g., Refs. \cite{lechner2018ClassicalElectrodynamicsModern,weinberg1995Foundations} for pedagogical introductions.
    
    What could eventual source terms indicate? First, notice that \(\dd{\mathcal{H}} = 0\) was enforced automatically due to the continuity equation. In fact, we could have defined 
    \begin{equation}\label{eq: H-form-nonperturbative}
        \mathcal{H} = \rho \dd{x} \wedge \dd{y} \wedge \dd{z} - \frac{1}{2} \epsilon_{ijk} (\rho v^i) \dd{t} \wedge \dd{x^j} \wedge \dd{x^k}
    \end{equation}
    and obtained \(\dd{\mathcal{H}} = 0\) as the nonperturbative expression of the mass-continuity equation. \(\dd\hodge \mathcal{H} = 0\), on the other hand, was obtained through the assumptions that the fluid flow is irrotational and barotropic. While this direction will not be pursued here, more general fluid flows involving vorticity or more intricate equations of motion could possibly be described by admitting a nonvanishing \(\dd\hodge \mathcal{H}\). Notice this would forbid a simple description in terms of the scalar potential \(\phi\), making the two-form formulation more interesting.
    
    Previous works had considered the idea of describing hydrodynamics in terms of a two-form \cite{nambu1977StringsVorticesGauge,sugamoto1979DualTransformationAbelian,matsuo2021NoteDescriptionPerfect,fischer1999MotionQuantizedVortices,lund1976UnifiedApproachStrings}. One of their main interests was the study of vorticity in fluid flow---two-form gauge fields naturally couple to one-dimensional objects, and vortices are naturally associated with one-dimensional vortex lines. These approaches do differ from ours, however, because our focus on linearized perturbations makes the theory naturally formulated on the ``acoustic spacetime'' introduced by \textcite{unruh1981ExperimentalBlackHoleEvaporation}, while previous analyses worked in Minkowski spacetime and included nonlinearities and vorticity as sources to the field equations. Most of the following analysis, however, will assume the background velocity to vanish. 

    \subsection{Gauge potential}\label{subsec: gauge-potential}
        The equations of motion for linear acoustics state that \(\mathcal{H}\) (a three-form) and \(\hodge \mathcal{H}\) (a one-form) are closed. In general, they do not need to be exact, and thus it may not be possible to obtain a two-form \(\mathcal{B}\) and/or a scalar field \(\phi\) such that \(\dd \mathcal{B} = \mathcal{H}\) and \( \dd\phi=\hodge \mathcal{H}\) (and, hence, \(\dd \mathcal{B} = \mathcal{H}  =\hodge\dd\phi\)  ). The conditions for a closed form to be exact are given by de Rham cohomology, which is briefly summarized in App. \ref{app: differential-forms} (see Refs. \cite{frankel2012GeometryPhysicsIntroduction,gorodski2020SmoothManifolds,lee2012IntroductionSmoothManifolds,nakahara2003GeometryTopologyPhysics,tu2011IntroductionManifolds} for various introductions). Note, however, that we are not interested in a theory formulated in a general manifold. We want to study sound in the radiation zone. This means we want to consider all angles around the source and all possible times, but we are only interested in large radii. For example, we can say we are only interested in what happens for \(r > R\), where all image sources are contained inside a sphere of radius \(R\). Topologically, this radiation zone is given by \(\mathbb{R}^2 \times \mathbb{S}^2\). The factor \(\mathbb{S}^2\) represents all possible angles to which radiation could go, one of the factors of \(\mathbb{R}\) represents time, and the remaining factor of \(\mathbb{R}\) is (topologically) the interval \((R,+\infty)\). In \(\mathbb{R}^2 \times \mathbb{S}^2\)---i.e., in the radiation zone---de Rham cohomology ensures that closed three-forms such as \(\mathcal{H}\) and closed one-forms such as \(\hodge \mathcal{H}\) are always exact. Hence, in the radiation zone it is always possible to introduce the two-form potential \(\mathcal{B}\) and the scalar potential \(\phi\) such that \(\dd \mathcal{B} = \mathcal{H} = \hodge \dd\phi\). \(\phi\) is precisely the scalar field we considered in Sec. \ref{sec: linear-acoustics} and that \textcite{unruh1981ExperimentalBlackHoleEvaporation} originally considered in analog gravity. It is the phonon field. Had we allowed vorticity in the fluid, we would not have \(\dd\hodge \mathcal{H} = 0\), and it would not be possible to introduce \(\phi\) even in the radiation zone. In the language of Sec. \ref{sec: linear-acoustics}, \(\curl{\vb{v}} \neq \vb{0}\) would prevent us from writing \(\vb{v} = \grad\phi\). We are now just formulating the same type of statement in the language of differential forms. 
         
         To pursue a description of asymptotic symmetries, it will be more convenient to work with \(\mathcal{B}\) rather than \(\phi\). In terms of \(\mathcal{B}\), linear acoustics becomes a gauge theory. The potential \(\mathcal{B}\) is not uniquely defined because we can perform transformations of the form 
        \begin{equation}
            \mathcal{B} \to \mathcal{B} + \mathcal{F},
        \end{equation}
        where \(\mathcal{F}\) is any closed two-form. If \(\mathcal{F}\) is closed, but not exact, the transformation is a global symmetry. In fact, the existence of a potential that is closed, but not exact, is the origin of the Aharonov--Bohm effect in electrodynamics \cite{sanders2014ElectromagnetismLocalCovariance}. If \(\mathcal{F}\) is exact, then the transformation is a gauge transformation. Gauge transformations that do not vanish at infinity will eventually become the asymptotic symmetries in this description.

        At this point, it is natural to ask whether closed \(p\)-forms in \(\mathbb{R}^2 \times \mathbb{S}^2\) are always exact. The answer is negative. Up to equivalence, there is precisely one two-form that is closed, but not exact. This is the volume form on the unit sphere (see App. \ref{app: differential-forms}). Since this form is closed, its effects are not captured by the field \(\mathcal{H}\), but they can still have physical manifestations. As previously mentioned, this is the same mechanism that gives rise to the Aharonov--Bohm effect in electrodynamics. Our investigations are not topological in nature, so the existence of this field configuration will not impact our results. Nevertheless, the existence of a closed, but nonexact, two-form on the sphere will be relevant for other reasons.

        An important subtlety of this theory---and of \(p\)-form electrodynamics more generally---is that it involves reducible gauge symmetries. Let us consider a gauge transformation for the \(\mathcal{B}\)-field. We have 
        \begin{equation}
            \mathcal{B} \to \mathcal{B} + \dd \mathcal{A},
        \end{equation}
        where \(\mathcal{A}\) is a one-form. This expression already ensures we are shifting \(\mathcal{B}\) by an exact form. Since a one-form has four-components at each spacetime point (assuming a four-dimensional spacetime), we may be tricked into believing we can eliminate four components of \(\mathcal{B}\). This, however, is not correct. Not all choices of \(\mathcal{A}\) are valid gauge transformations. If \(\mathcal{A}\) happens to be exact (\(\mathcal{A} = \dd\lambda\)), then the alleged gauge transformation would be 
        \begin{equation}
            \mathcal{B} \to \mathcal{B} + \dd[2]\lambda = \mathcal{B}.
        \end{equation}
        Hence, the gauge transformation becomes trivial. The issue is that \(\mathcal{A}\) itself has gauge symmetries. In the jargon, there are gauge-for-gauge symmetries, or the gauge symmetry is reducible. \(p\)-form electrodynamics is, in fact, the textbook example of a theory with reducible gauge symmetry \cite{henneaux1990LecturesAntifieldBRSTFormalism,henneaux1992QuantizationGaugeSystems,weinberg1996ModernApplications,esmaeili2020pFormGaugeFields}. Notice this does not occur in Maxwell electrodynamics because \(0\)-forms are only exact if they vanish.

        This subtlety means that, to truly fix the gauge of the theory, we also need to fix the gauge of the gauge parameters. As an example, suppose we are interested in Lorenz gauge, which is employed in the analyses by \textcite{campiglia2019ScalarAsymptoticCharges,manzoni2026HigherorderpformAsymptotic,francia2024AsymptoticChargespforms}, for instance. Noticing the gauge-for-gauge symmetry
        \begin{equation}
            \mathcal{A} \to \mathcal{A}' = \mathcal{A} + \dd\lambda,
        \end{equation}
        we choose \(\lambda\) to satisfy
        \begin{equation}
            \dd\hodge\dd\lambda = - \dd\hodge \mathcal{A},
        \end{equation}
        which is the wave equation for \(\lambda\) with a source. Solving this equation, we can enforce \(\dd\hodge \mathcal{A}' = 0\) (and have thus reached Lorenz gauge for \(\mathcal{A}\)), and are left with a residual gauge-for-gauge symmetry in \(\lambda\). Now that \(\mathcal{A}\) has been gauge-fixed, we notice that
        \begin{equation}
            \mathcal{B} \to \mathcal{B}' = \mathcal{B} + \dd \mathcal{A},
        \end{equation}
        and thus choose to solve
        \begin{equation}
            \dd\hodge\dd \mathcal{A} = - \dd\hodge \mathcal{B}.
        \end{equation}
        Given that \(\dd\hodge \mathcal{A} = 0\) from the previous step, we see we are trying to solve Maxwell's equations with a source current \(- \dd\hodge \mathcal{B}\). This source is conserved because \(\dd[2] = 0\). Hence, we can enforce \(\dd\hodge \mathcal{B}' = 0\), and thus reach Lorenz gauge for \(\mathcal{B}\), with a residual gauge symmetry in \(\mathcal{A}\). 

        The outcome is that we now have the equations
        \begin{subequations}
            \begin{align}
                \dd\hodge\dd \mathcal{B} &= 0, \\
                \dd\hodge\dd \mathcal{A} &= 0, \\
                \dd\hodge\dd \lambda &= 0,
            \end{align}
        \end{subequations}
        where \(\mathcal{A}\) and \(\mathcal{B}\) are assumed to be in Lorenz gauge. The first equation is the equation of motion for the theory, while the remaining equations give the conditions on residual gauge symmetry.

        Originally, \(\mathcal{B}\) seemed to have six independent components, \(\mathcal{A}\) seemed to have four, and \(\lambda\) had one. Once \(\mathcal{A}\) was gauge-fixed, it had only three independent components. We used these three components to eliminate three components of \(\mathcal{B}\). At this point, \(\mathcal{B}\) has 
        \begin{equation}
            6 - (4 - 1) = 3
        \end{equation}
        independent components, where the parentheses compute the number of independent components of \(\mathcal{A}\) taking gauge-for-gauge symmetry into consideration. Before using residual gauge symmetry, we thus have three components in \(\mathcal{B}\), three in \(\mathcal{A}\), and one in \(\lambda\). The residual freedom in \(\lambda\) lets us reduce the components of \(\mathcal{A}\) from three to two. These two components then allow us to reduce the independent components of \(\mathcal{B}\) from three to one. Hence, the number of degrees of freedom in \(\mathcal{B}\) is
        \begin{equation}
            3 - (3 - 1) = 1,
        \end{equation}
        which is expected from the duality \(\dd\mathcal{B} = \hodge\dd\phi\), since \(\phi\) also carries a single physical degree of freedom. See also Refs. \cite{henneaux1992QuantizationGaugeSystems,esmaeili2020pFormGaugeFields} for other ways of doing this counting of number of degrees of freedom.

    \subsection{Image sources}
        At this stage, one could worry about the source term \(q\) that we introduced in Sec. \ref{sec: linear-acoustics}. Nonperturbatively, such a source term would affect the continuity equation so that \(\dd{\mathcal{H}} = \hodge q\). Hence, \(\mathcal{H}\) is no longer closed, which spoils the introduction of the gauge field \(\mathcal{B}\). This is not a problem. Recall that the source \(q\) was introduced in a nonphysical region to model the boundary conditions imposed on the acoustic perturbations. What this source means now is that we are not interested in working with \(\mathbb{R}^4\) as a manifold, but rather with the radiation zone \(\mathbb{R}^2 \times \mathbb{S}^2\), which removes a compact set from the center of space. In the radiation zone, which is the physical region we are interested in, \(\mathcal{H}\) is a closed form. When trying to extend \(\mathcal{H}\) to \(\mathbb{R}^4\) we are forced to introduce an image source term.
    
        As an explicit example, let us consider \(\mathcal{H}\) defined on \(\mathbb{R}^4\), so that \(\dd \mathcal{H} = \hodge q\). Then we can integrate over a timelike cylinder \(\mathcal{V}\) defined by all points in spacetime with \(r < R\). Notice the cylinder is timelike, and we ignore any spacelike ``caps.'' We find
        \begin{subequations}
            \begin{align}
                \int_{\mathcal{V}} \hodge q &= \int_{\partial\mathcal{V}} \mathcal{H}, \\
                &= \int_{-\infty}^{+\infty}\oint_{\mathbb{S}^2} \tensor{\mathcal{H}}{_\theta_\varphi_t} \dd{\theta} \wedge \dd{\varphi} \wedge \dd{t}, \\
                &= - \int_{-\infty}^{+\infty}\oint_{\mathbb{S}^2} \rho \vb{v} \vdot \vu{r} \, R^2 \sin{\theta} \dd{\theta} \dd{\varphi} \dd{t}.
            \end{align}
        \end{subequations}
        The final integrand obeys the conservation equation away from the source (i.e., it obeys the conservation equation in a region with topology \(\mathbb{R}^2\times\mathbb{S}^2\)). If we want to write the surface integral as a volume integral, however, we would need to introduce a source term. This would spoil the gauge-theoretic description, but it is essentially the method of images we used in Sec. \ref{sec: linear-acoustics}.

\section{Case Study: The Maxwell Infrared Triangle}\label{sec: electrodynamics}
    Before working out the acoustic case in detail, let us first discuss Maxwell electrodynamics. This will serve to illustrate most of the techniques we will employ and allow us to set notation and conventions relative to most of the literature. Once electrodynamics is clear, acoustics will follow naturally. Below, we adapt discussions by \textcite{bieri2013ElectromagneticAnalogGravitational,pasterski2017AsymptoticSymmetriesElectromagnetic,satishchandran2019AsymptoticBehaviorMassless,bieri2024ExperimentMeasureElectromagnetic}. We work in Heaviside--Lorentz units with the speed of light set to \(c = 1\), so the Maxwell equations are written as \cite{misner2017Gravitation}
    \begin{equation}
        \dd\hodge\mathcal{F} = \hodge j,
    \end{equation}
    where \(j\) is the current one-form and \(\mathcal{F}\) is the Faraday two-form.

    \subsection{The memory effect}
        The Lorentz force law is given by 
        \begin{equation}
            \tensor{u}{^b}\tensor{\nabla}{_b}\tensor{u}{^a} = \frac{q}{m}\tensor{\mathcal{F}}{^a_b}\tensor{u}{^b},
        \end{equation}
        where \(\tensor{\mathcal{F}}{_a_b}\) is the Faraday tensor, \(\tensor{u}{^a}\) is the four-velocity of a test particle, \(q\) is the particle's electric charge, and \(m\) is its mass. Let \(\lambda\) be an affine parameter along the integral lines of \(\tensor{u}{^a}\). Then
        \begin{equation}
            \tensor{u}{^a}(\lambda) = \tensor*{u}{^a_0} + \frac{q}{m} \int_{-\infty}^{\lambda} \tensor{\mathcal{F}}{^a_b}\tensor{u}{^b} \dd{\lambda'},
        \end{equation}
        where \(\tensor*{u}{^a_0} \equiv \tensor*{u}{^a}(-\infty)\) is the value of the four-velocity at early times. 
        
        To ensure the spacetime has finite total charge, it is necessary that \(\tensor{\mathcal{F}}{_a_b}\) falls off at least as \(\order{1/r}\) for \(r \to +\infty\). As a consequence, the difference \(\tensor{u}{^a}(\lambda) - \tensor*{u}{^a_0}\) also has to fall off at least as \(\order{1/r}\). We then see that 
        \begin{equation}
            \tensor{u}{_a}(\lambda) = \tensor*{u}{_a^0} + \frac{q}{m r} \int_{-\infty}^{\lambda} \tensor*{\mathcal{F}}{^{(1)}_a_b} \dd{\lambda'} \tensor*{u}{^b_0} + o\qty(\frac{1}{r}),
        \end{equation}
        where
        \begin{equation}
            \tensor*{\mathcal{F}}{^{(1)}_a_b} = \lim_{r \to +\infty} r \tensor{\mathcal{F}}{_a_b}.
        \end{equation}
        This limit must be well-defined for a finite-energy configuration.
        
        Taking \(\lambda \to +\infty\), we conclude that 
        \begin{equation}
            \Delta\tensor{u}{_a} = \frac{q}{m r} \int_{-\infty}^{+\infty} \tensor*{\mathcal{F}}{^{(1)}_a_b} \dd{\lambda'} \tensor*{u}{^b_0} + o\qty(\frac{1}{r}),
        \end{equation}
        where we introduced the shorthand
        \begin{equation}
            \Delta \tensor{u}{^a} = \tensor{u}{^a}(+\infty) - \tensor{u}{^a}(-\infty).
        \end{equation}
        Hence, after the passage of an electromagnetic wave, a test charge will undergo a velocity kick. This kick is directly related to a nonvanishing integral of the field-strength tensor. We can then define the ``memory tensor'' \(\tensor{\Delta}{_a}\) as 
        \begin{equation}
            \tensor{\Delta}{_a} = \int_{-\infty}^{+\infty} \tensor*{\mathcal{F}}{^{(1)}_a_b} \dd{\lambda'} \tensor*{u}{^b_0}.
        \end{equation}
        We are particularly interested in the behavior near null infinity. Hence, let us introduce Bondi coordinates \cite{bondi1960GravitationalWavesGeneral} and write the Minkowski line element as 
        \begin{equation}\label{eq: Bondi-coordinates-Minkowski}
            \dd{s}^2 = - \dd{u}^2 - 2 \dd{u} \dd{r} + r^2 \tensor{\gamma}{_A_B}\dd{\tensor{x}{^A}}\dd{\tensor{x}{^B}}.
        \end{equation}
        Above, \(u\) is the retarded time 
        \begin{equation}
            u = t - r,
        \end{equation}
        while \(\tensor{\gamma}{_A_B}\) is the round metric on the sphere. Capital indices are always raised and lowered with \(\tensor{\gamma}{_A_B}\). The Levi-Civita connection for \(\tensor{\gamma}{_A_B}\) will be denoted \(\tensor{\mathcal{D}}{_A}\). We also write \(\mathcal{D}^2 = \tensor{\mathcal{D}}{^A}\tensor{\mathcal{D}}{_A}\) for the Laplacian on the sphere.
        
        We then consider the four-velocity 
        \begin{equation}
            \tensor*{u}{^a_0} = \tensor{\qty(\pdv{u})}{^a}.
        \end{equation}
        Notice that, in Bondi coordinates, this expression is indeed the four-velocity of a timelike inertial observer. Also notice \(\tensor*{u}{^a}(\lambda)\) will not generically be given by this expression, because it is not in inertial motion---rather, its acceleration is dictated by the Lorentz force law. We conclude then that a sensible definition of ``memory'' for the electromagnetic field is the tensor 
        \begin{equation}
            \tensor{\Delta}{_a} = \int_{-\infty}^{+\infty} \tensor*{\mathcal{F}}{^{(1)}_a_u} \dd{u}.
        \end{equation}
        We say there is electromagnetic memory whenever \(\tensor{\Delta}{_a}\) is nonvanishing. This will be physically measurable in terms of \(\Delta \tensor{u}{^a}\). Furthermore, notice the memory tensor is defined in a gauge-invariant manner. 
    
        It is convenient to express these quantities in the Bondi coordinate system. Earlier, we defined \(\tensor*{\mathcal{F}}{^{(1)}_a_b}\) to be the \(1/r\) component of the \emph{tensor} \(\tensor{\mathcal{F}}{_a_b}\). This is the convention used, for example, by \textcite{satishchandran2019AsymptoticBehaviorMassless}. It is convenient, however, to see the decay properties of each of the components of \(\tensor{\mathcal{F}}{_a_b}\) in the Bondi coordinate basis. We are assuming the field components to fall off at large \(r\) (and constant \(u\)) as
        \begin{subequations}\label{eq: falloffs-electromagnetism}
            \begin{align}
                \tensor{\mathcal{F}}{_u_r} &= \frac{\tensor*{F}{_u_r}}{r^2} + o\qty(\frac{1}{r^2}), \\
                \tensor{\mathcal{F}}{_A_r} &= \frac{\tensor*{F}{_A_r}}{r^2} + o\qty(\frac{1}{r^2}), \\
                \tensor{\mathcal{F}}{_A_u} &= \tensor*{F}{_A_u} + o\qty(1), \\
                \tensor{\mathcal{F}}{_A_B} &= \tensor*{F}{_A_B} + o\qty(1).
            \end{align}
        \end{subequations}
        These conditions are loose enough to be interesting, but strong enough to ensure a finite-energy (and finite-charge) configuration. Note we are giving the expressions are in a coordinate basis. The \(1/r\) behavior of the \emph{tensor} should be defined in an orthonormal basis. In this case one has, for example,
        \begin{equation}
            \frac{1}{r}\tensor{\qty(\pdv{\tensor{x}{^A}})}{^a}\tensor{\qty(\pdv{u})}{^b}\tensor{\mathcal{F}}{_a_b} = \frac{\tensor*{F}{_A_u}}{r} + o\qty(\frac{1}{r}).
        \end{equation}
        Hence, 
        \begin{equation}
            \tensor*{\mathcal{F}}{^{(1)}_A_u} = \tensor{F}{_A_u}.
        \end{equation}
    
        With this in mind, we can see the memory tensor only has angular components, because \(\tensor{\mathcal{F}}{_u_u} = 0\) due to antisymmetry and \(\tensor*{\mathcal{F}}{^{(1)}_u_r} = 0\) because the electromagnetic radiative field is transversal. Hence, we can safely write
        \begin{equation}\label{eq: definition-maxwell-memory-tensor}
            \tensor{\Delta}{_A} = \int_{-\infty}^{+\infty} \tensor{F}{_A_u} \dd{u}
        \end{equation}
        and know we are accounting for all memory components. Note then that the memory tensor is a one-form on the sphere. Since we are integrating over \(u\) and picking a specific coefficient in a large-\(r\) expansion, the memory tensor does not depend on either \(r\) or \(u\). 
    
        We can relate the memory tensor to changes in the Coulombic field and to radiation piercing null infinity. To do so, consider the Maxwell equation\footnote{The Supplemental Material accompanying this preprint includes a \textsc{Mathematica} \cite{wolframresearch2026Mathematica150} notebook used to rewrite several equations of motion in Bondi coordinates. This includes the wave equation, the Maxwell equations, and some other expresions used throughout the paper. The code is written in \textsc{xAct} \cite{martin-garciaXActEfficientTensor,martin-garcia2008XPermFastIndex} and is based on an example notebook by \textcite{stein2018CalculateCCESystem}.}
        \begin{equation}
            -\tensor{\partial}{_u}\tensor{\mathcal{F}}{_u_r} + \frac{1}{r^2}\pdv{r}\qty(r^2 \tensor{\mathcal{F}}{_u_r}) - \frac{\tensor{\mathcal{D}}{^A}\tensor{\mathcal{F}}{_A_u}}{r^2} = \tensor{j}{_u},
        \end{equation}
        which is essentially Gauss' law in Bondi coordinates. The leading component in a large-\(r\) expansion reads 
        \begin{equation}
            -\tensor{\partial}{_u}\tensor*{F}{_u_r} - \tensor{\mathcal{D}}{^A}\tensor*{F}{_A_u} = \tensor*{j}{^{(2)}_u},
        \end{equation}
        where \(\tensor*{j}{^{(2)}_u}\) is the \(1/r^2\) component of \(\tensor*{j}{_u}\).
        If we integrate over \(u\) we find
        \begin{equation}\label{eq: ordinary-null-memory}
            \tensor{\mathcal{D}}{^A}\tensor{\Delta}{_A} = - \Delta\tensor*{F}{_u_r} - \int \tensor*{j}{^{(2)}_u} \dd{u}.
        \end{equation}
        We thus see the memory tensor is associated to a shift in the Coulombic field---known as ordinary memory \cite{bieri2013ElectromagneticAnalogGravitational}---and to the escape of electric current through null infinity---known as null memory \cite{bieri2013ElectromagneticAnalogGravitational}.
    
        Because \(\mathcal{F}\) is a closed form (\(\dd \mathcal{F} = 0\)), its components obey the equation
        \begin{equation}
            \tensor{\partial}{_u}\tensor{\mathcal{F}}{_A_B} = 2\tensor{\mathcal{D}}{_[_B}\tensor{\mathcal{F}}{_A_]_u},
        \end{equation}
        where \(\tensor{\mathcal{D}}{_A}\) is the Levi-Civita connection on the unit sphere. This is essentially Faraday's law written in Bondi coordinates and covariant notation. If we integrate both sides, we find (at leading order in a large-\(r\) expansion)
        \begin{equation}
            \left.\tensor*{F}{_A_B}\right|_{+\infty} - \left.\tensor*{F}{_A_B}\right|_{-\infty} = 2\tensor{\mathcal{D}}{_[_B}\tensor{\Delta}{_A_]}.
        \end{equation}
        The left-hand side poses an obstruction for the memory tensor to be a closed one-form. We will assume---as done by \textcite{bieri2013ElectromagneticAnalogGravitational,pasterski2017AsymptoticSymmetriesElectromagnetic}---that \(\tensor*{F}{_A_B}\) vanishes for \(u \to \pm \infty\). This is indeed the case for the Liénard--Wiechert solution, and thus for an arbitrary superposition of massive electric charges. Massless charges are still allowed in the spacetime, as long as they are constrained to finite values of \(u\) (which is often the case of interest). With this assumption, we conclude that
        \begin{equation}
            2\tensor{\mathcal{D}}{_[_A}\tensor{\Delta}{_B_]} = 0,
        \end{equation}
        meaning \(\tensor{\Delta}{_A}\) is a closed one-form on the sphere. De Rham cohomology then tells us that \(\tensor{\Delta}{_A}\) is necessarily exact. Therefore, there must exist a function \(\varepsilon\) on the sphere such that 
        \begin{equation}\label{eq: memory-1-form-exact}
            \tensor{\mathcal{D}}{_A}\varepsilon = \tensor{\Delta}{_A}.
        \end{equation}
    
        If we rewrite Eq. \eqref{eq: ordinary-null-memory} in terms of \(\varepsilon\), we find
        \begin{equation}\label{eq: large-gauge-from-memory}
            \mathcal{D}^2\varepsilon = - \Delta\tensor*{F}{_u_r} - \int \tensor*{j}{^{(2)}_u} \dd{u},
        \end{equation}
        where \(\mathcal{D}^2 = \tensor{\mathcal{D}}{^A}\tensor{\mathcal{D}}{_A}\) is the Laplacian on the sphere. This is the Poisson equation on the sphere. It can be solved as long as the right-hand side has vanishing average on the sphere. This is the case, because we already know the right-hand side is the divergence of a vector field. The Laplacian \(\mathcal{D}^2\) can then be inverted and \(\varepsilon\) is uniquely defined.

    \subsection{The asymptotic symmetries}
        Let us now consider electrodynamics in terms of the four-potential. Generically, we have 
        \begin{equation}
            \mathcal{F} = \dd\mathcal{A}.
        \end{equation}
        We can always choose to work in Lorenz gauge, in which case the four-potential has to obey
        \begin{equation}
            \dd\hodge\mathcal{A} = 0.
        \end{equation}
        Within Lorenz gauge, there is a residual gauge symmetry given by 
        \begin{equation}
            \mathcal{A} \to \mathcal{A} + \dd\lambda,
        \end{equation}
        where \(\lambda\) is required to satisfy
        \begin{equation}
            \dd\hodge\dd\lambda = 0.
        \end{equation}
        This is the wave equation for \(\lambda\).

        We are not interested in arbitrary field configurations, but rather in configurations with the falloffs \eqref{eq: falloffs-electromagnetism}. To achieve them, we impose that the potential obeys
        \begin{subequations}\label{eq: falloffs-A-field}
            \begin{align}
                \tensor{\mathcal{A}}{_u} &= \frac{\tensor{A}{_u}}{r} + \tensor{\partial}{_u}\lambda + o\qty(\frac{1}{r}), \\
                \tensor{\mathcal{A}}{_r} &= \frac{\tensor{A}{_r}}{r^2} + \tensor{\partial}{_r}\lambda + o\qty(\frac{1}{r^2}), \\
                \tensor{\mathcal{A}}{_A} &= \tensor{A}{_A} + \tensor{\mathcal{D}}{_A}\lambda + o\qty(1),
            \end{align}
        \end{subequations}
        where \(\lambda\) is a scalar function---a gauge parameter---with the expansion
        \begin{equation}\label{eq: polyhomogeneous-lambda}
            \lambda(u,r,\tensor{x}{^A}) \sim \sum_{k=0} \frac{\lambda^{(k)}}{r^k} + \sum_{k=1} \frac{\hat{\lambda}^{(k)} \log{r}}{r^k}.
        \end{equation}
        These choices of falloffs impose boundary conditions on the configurations we are interested in. They define what is the phase space for the theory. See, for example, Refs. \cite{compere2019AdvancedLecturesGeneral,ruzziconi2020AsymptoticSymmetriesGauge}. The choice to include looser falloffs in the pure gauge section (i.e., the choice of allowing \(\lambda\) to enter the falloffs for \(\mathcal{A}\)) is a necessity of Lorenz gauge \cite{francia2024AsymptoticChargespforms,campoleoni2019ElectromagneticColourMemory,satishchandran2019AsymptoticBehaviorMassless}. An analysis in radial gauge---which was preferred in Refs. \cite{he2014NewSymmetriesMassless,strominger2018LecturesInfraredStructure,pasterski2017AsymptoticSymmetriesElectromagnetic}, for example---would not need logarithmic contributions, but it would lead to difficulties when generalizing to a two-form theory \cite{francia2018TwoFormAsymptoticSymmetries}.

        In Bondi coordinates, the wave equation satisfied by \(\lambda\) can be written as 
        \begin{equation}
            -\frac{2}{r}\pdv{r}\qty(r \tensor{\partial}{_u}\lambda) + \frac{1}{r^2}\pdv{r}\qty(r^2 \tensor{\partial}{_r}\lambda) + \frac{\mathcal{D}^2 \lambda}{r^2} = 0.
        \end{equation}
        In terms of the asymptotic expansion, we obtain\footnote{The Supplemental Material accompanying this preprint includes a \textsc{Mathematica} \cite{wolframresearch2026Mathematica150} notebook used to compute the large-\(r\) expansion of the wave equation (and other relevant equations) in Bondi coordinates.}
        \begin{subequations}
            \begin{multline}
                2(n-1)\tensor{\partial}{_u}\lambda^{(n)} - 2 \tensor{\partial}{_u}\hat{\lambda}^{(n)} - (2n-3)\hat{\lambda}^{(n-1)} \\ + (\mathcal{D}^2 + (n-1)(n-2))\lambda^{(n-1)} = 0
            \end{multline}
            and
            \begin{equation}
                2(n-1)\tensor{\partial}{_u}\hat{\lambda}^{(n)} + (\mathcal{D}^2 + (n-1)(n-2))\hat{\lambda}^{(n-1)} = 0.
            \end{equation}
        \end{subequations}
        For \(n=0\), we find
        \begin{equation}
            \tensor{\partial}{_u}\lambda^{(0)} = 0,
        \end{equation}
        and thus the leading term is time-independent. For \(n=1\), we find
        \begin{equation}
            2 \tensor{\partial}{_u}\hat{\lambda}^{(1)} = \mathcal{D}^2\lambda^{(0)},
        \end{equation}
        which determines \(\hat{\lambda}^{(1)}\) from \(\lambda^{(0)}\) and an initial condition \(\hat{\lambda}^{(1)}(-\infty,\tensor{x}{^A})\). Notice \(\tensor{\partial}{_u}\lambda^{(1)}\) is not constrained by the equations, and thus can be specified arbitrarily\footnote{When solving the wave equation at null infinity, the freedom of specifying \(\lambda^{(1)}\) arbitrarily corresponds to picking what radiation reaches null infinity. This term would be constrained if we had solved the equations in the bulk of spacetime and only then expanded near null infinity. Nevertheless, by solving the equations directly near null infinity, we cannot probe the deep interior of the bulk, and thus must now inform its effects by specifying the radiation it emitted in the form of \(\lambda^{(1)}\).}. For \(n > 1\), the equations can be solved hierarchically in terms of initial data for each \(\lambda^{(n)}\) and \(\hat{\lambda}^{(n)}\) and knowledge of the previous coefficients. 

        Most importantly, we notice the existence of residual gauge transformations---i.e., gauge transformations preserving Lorenz gauge---that do not vanish at infinity. These transformations have the asymptotic behavior
        \begin{equation}
            \lambda(u,r,\tensor{x}{^A}) = \varepsilon(\tensor{x}{^A}) + \order{\frac{\log{r}}{r}},
        \end{equation}
        where we wrote \(\lambda^{(0)}(\tensor{x}{^A}) = \varepsilon(\tensor{x}{^A})\) to simplify the notation. Since \(\varepsilon(\tensor{x}{^A})\) remains finite at infinity, these are known as large gauge transformations.

        Now let us connect this discussion to memory. Because the memory tensor \(\tensor{\Delta}{_A}\) is an exact form, it can be understood as a gauge transformation. To make this concrete, let us write it in terms of the four-potential. In any globally defined gauge, we can write
        \begin{equation}
            \tensor{\mathcal{F}}{_A_u} = \tensor{\mathcal{D}}{_A}\tensor{\mathcal{A}}{_u} - \tensor{\partial}{_u}\tensor{\mathcal{A}}{_A}.
        \end{equation}
        At leading order, 
        \begin{equation}
            \tensor{F}{_A_u} = - \tensor{\partial}{_u}\tensor{A}{_A},
        \end{equation}
        where we assume \(\mathcal{A}\) to fall off as in Eq. \eqref{eq: falloffs-A-field}. Notice the memory tensor is given by \begin{equation}
            \tensor{\Delta}{_A} = - \Delta\tensor{A}{_A}.
        \end{equation}
        If we recall Eq. \eqref{eq: memory-1-form-exact}, we conclude that 
        \begin{equation}\label{eq: large-gauge-from-memory-A}
            \Delta\tensor*{A}{_A} = -\tensor{\mathcal{D}}{_A}\varepsilon.
        \end{equation}
        In other words, the difference between the late and early time values of the potential \(\tensor*{A}{^{(1)}_A}\) is an exact form on the sphere. 

        What is the physical meaning of this? Suppose that at early times (\(u \to -\infty\)) there is no radiation in the spacetime. At this time, we choose a gauge and define that the vacuum---the absence of electromagnetic radiation---is to be understood as \(\tensor{\mathcal{A}}{_A} = 0\).

        During intermediate times, radiation is emitted to infinity. For example, charges get accelerated and radiate. These bursts of radiation will generically change the value of the Coulombic field, because the final configuration will involve charges with different velocities than the initial configuration. At late times, we are once again in the absence of radiation. However, due to the memory effect, we no longer have \(\tensor{\mathcal{A}}{_A} = 0\). Instead, the potential for a vacuum (i.e., radiation-free) configuration is now a gauge transformation away, in accordance with Eq. \eqref{eq: large-gauge-from-memory-A}. The definition of what is a vacuum has changed. The natural Fock spaces for quantum electrodynamics at early and late times are also inequivalent---see Refs. \cite{satishchandran2019AsymptoticBehaviorMassless,prabhu2022InfraredFiniteScattering} for thorough discussions.

        Importantly, the memory tensor can be written as a large gauge transformations. More specifically, we can always write 
        \begin{equation}
            \tensor{\Delta}{_\mu} = \tensor{\partial}{_\mu}\lambda + o(1).
        \end{equation}
        This is true because \(\tensor{\Delta}{_u} = \tensor{\Delta}{_r}\) will vanish, but the leading-order behavior of \(\lambda\) is independent of both \(u\) and \(r\). Hence, at leading order, the memory in the electromagnetic field is given precisely by a large gauge transformation. Equation \eqref{eq: large-gauge-from-memory} tells us how to choose a large gauge transformation (based on its leading behavior \(\varepsilon\)) that ``resets'' the potential to zero at late times.

    \subsection{The soft theorem}
        Soft theorems are arguably the best understood corner of the infrared triangle, and thus we will merely sketch their relation to the rest of the triangle. 

        Using the covariant phase space formalism \cite{lee1990LocalSymmetriesConstraints,iyer1994PropertiesNoetherCharge,iyer1995ComparisonNoetherCharge,wald2000GeneralDefinitionConserved,barnich2002CovariantTheoryAsymptotic,gieres2023CovariantCanonicalFormulations,compere2019AdvancedLecturesGeneral,fiorucci2021LeakyCovariantPhase}, we could define the large gauge Noether charges
        \begin{equation}
            Q_+(\varepsilon) = \oint_{\mathscr{I}^+_-} \varepsilon \wedge \hodge\mathcal{F}.
        \end{equation}
        We denote by \(\mathscr{I}^+_-\) the ``past boundary'' of future null infinity (understood as the limit \(u \to - \infty\)). Similarly, we can define charges in past null infinity, 
        \begin{equation}
            Q_-(\varepsilon) = \oint_{\mathscr{I}^-_+} \varepsilon \wedge \hodge\mathcal{F}, 
        \end{equation}
        where \(\mathscr{I}^-_+\) is the ``future boundary'' of past null infinity (in the sense of the limit \(v \to + \infty\), with \(v\) the advanced time). We could then obtain the Poisson brackets for these charges, and use them to obtain the commutators they need to respect in the quantized theory. Imposing that the quantum charges commute with the \(S\)-matrix then leads to 
        \begin{equation}\label{eq: ward-identity}
            \mel{\text{out}}{Q_+(\varepsilon) S - S Q_-(\varepsilon)}{\text{in}} = 0,
        \end{equation}
        for arbitrary in and out states. Equation~\eqref{eq: ward-identity} constrains different scattering amplitudes in the theory, and is known (in this particular case) as the Weinberg soft photon theorem \cite{he2014NewSymmetriesMassless,campiglia2015AsymptoticSymmetriesQED,weinberg1965InfraredPhotonsGravitons}. See, for example, Refs. \cite{strominger2018LecturesInfraredStructure,aguiaralves2026LecturesBondiMetzner} for a detailed discussion.

        In turn, it is possible to use quantum field theoretic scattering amplitudes to obtain information about classical observables \cite{kosower2019AmplitudesObservablesClassical,cristofoli2022WaveformsAmplitudes,kosower2022SAGEXReviewScattering,mohanty2023GravitationalWavesQuantum}. The soft theorem gives information about the low-energy behavior of photons, which, when Fourier transformed, is mapped into a long-distance behavior, recovering the memory effect. See also the discussions in Refs. \cite{satishchandran2019AsymptoticBehaviorMassless,prabhu2022InfraredFiniteScattering}.

\section{Memory in the Two-Form Formalism}\label{sec: memory-kalb-ramond}
    Armed with a thorough understanding of the Maxwell infrared triangle, let us now move back to linear acoustics in the gauge-theoretic formulation. We will derive the memory effect in terms of the two-form potential, and then move on to relate it to asymptotic symmetries. 

    Memory effects can be classified as linear, nonlinear, ordinary, and null \cite{bieri2014PerturbativeGaugeInvariant,bieri2024GravitationalWaveDisplacement,satishchandran2019AsymptoticBehaviorMassless}. ``Linear'' and ``nonlinear'' refer to the original calculations in general relativity, in which the memory effect was first identified in a linearized analysis \cite{zeldovich1974RadiationGravitationalWaves,braginsky1985KinematicResonanceMemory,braginsky1987GravitationalwaveBurstsMemory} and a nonlinear counterpart was understood much later \cite{christodoulou1991NonlinearNatureGravitation,blanchet1992HereditaryEffectsGravitational,thorne1992GravitationalwaveBurstsMemory}. ``Ordinary'' and ``null'' correspond to the two pieces given in Eq. \eqref{eq: ordinary-null-memory}. As highlighted by \textcite{bieri2013ElectromagneticAnalogGravitational,bieri2014PerturbativeGaugeInvariant}, the null contribution is due to massless fields that reach infinity. 

    In order to discuss null memory in acoustics, we would need to include sources in the equations of motion that can reach null infinity. On the one hand, this is physically reasonable. In fact, in acoustics, it is very reasonable to consider supersonic sources, which could lead to a type of ``supersonic memory,'' possibly similar to the effects described by \textcite{heisenberg2025ConstrainingSuperluminalEinsteinAEther,zosso2025EnhancementElectromagneticMemory}. We will not pursue this route here, both for simplicity and because such an analysis seems more natural once nonlinear effects in fluid flow are also being taken into account. As a consequence of our linear approximations, we then ignore any sorts of null memory, and focus instead on the case of (linear) ordinary memory. This will already allow us to investigate connections to the other corners of an acoustic infrared triangle.

    From this point onward, we settle for simplicity and take the background medium to be homogeneous and quiescent, which means the acoustic metric \eqref{eq: acoustic-metric} is the Minkowski metric up to coordinate redefinitions. This assumption seems analogous to the choice of working with Maxwell electrodynamics in a flat background, as opposed to in a curved spacetime. We choose units with \(\rho_0 = c = 1\), meaning the problem has effectively been reduced to studying the dynamics of a two-form field in a flat spacetime. As mentioned in the introduction, aspects of this problem were investigated by other authors with various motivations \cite{divecchia2015SoftTheoremGraviton,donnay2023pFormsCelestialSphere,afshar2018AsymptoticSymmetriespForm,afshar2019StringMemoryEffect,esmaeili2020pFormGaugeFields,francia2024AsymptoticChargespforms,heissenberg2019TopicsAsymptoticSymmetries,manzoni2026HigherorderpformAsymptotic,manzoni2026DualityAsymptoticCharges,romoli2024OrNTwoformAsymptotic}.

    As in the electromagnetic case, we equip flat spacetime with Bondi coordinates, which brings the Minkowski metric into the form \eqref{eq: Bondi-coordinates-Minkowski}. In these coordinates, the components of the three-form \(\mathcal{H}\) are 
    \begin{subequations}
        \begin{align}
            \tensor{\mathcal{H}}{_u_j_k} &= - \tensor{\epsilon}{_i_j_k} \tensor{\var v}{^i}, \\
            \tensor{\mathcal{H}}{_i_j_k} &= \tensor{\epsilon}{_i_j_k} (\var \rho - \tensor{\var v}{^r}),
        \end{align}
    \end{subequations}
    where the indices \(i, j, k, \ldots\) run over the coordinates in the \(u=\text{constant}\) manifolds. \(\tensor{\epsilon}{_i_j_k}\) denotes the volume element on these hypersurfaces, with
    \begin{equation}
        \tensor{\epsilon}{_r_\theta_\varphi} = r^2 \sin\theta
    \end{equation}
    in standard spherical coordinates. \(\tensor{\var v}{^i}\) denotes components in a coordinate basis, not in an orthonormal basis. For example,
    \begin{equation}
        \var\vb{v}\vdot\vu*{\varphi} = r \sin\theta \tensor{\var v}{^\varphi}.
    \end{equation}

    It is also convenient to write the components of \(\mathcal{H}\) in terms of the scalar field \(\phi\). We find
    \begin{subequations}
        \begin{align}
            \tensor{\mathcal{H}}{_u_A_B} &= r^2 \tensor{\epsilon}{_A_B} (\tensor{\partial}{_u}\phi - \tensor{\partial}{_r}\phi), \label{eq: HuAB-from-phi} \\
            \tensor{\mathcal{H}}{_r_A_B} &= - r^2 \tensor{\epsilon}{_A_B} \tensor{\partial}{_r}\phi, \\
            \tensor{\mathcal{H}}{_u_r_A} &= - \tensor{\epsilon}{_A_B} \tensor{\gamma}{^B^C}\tensor{\partial}{_C}\phi,
        \end{align}
    \end{subequations}
    where \(\tensor{\epsilon}{_A_B}\) is the volume form on the unit sphere. 

    Given our previous calculations with \(\phi\), we see we can consider the falloffs
    \begin{subequations}\label{eq: falloffs-H}
        \begin{align}
            \tensor{\mathcal{H}}{_u_A_B} &= r \tensor{H}{_u_A_B} + o\qty(r), \\
            \tensor{\mathcal{H}}{_r_A_B} &= \tensor{H}{_r_A_B} + o\qty(1), \\
            \tensor{\mathcal{H}}{_u_r_A} &= \frac{\tensor{H}{_u_r_A}}{r} + o\qty(\frac{1}{r}).
        \end{align}
    \end{subequations}
    These same falloffs could be argued for by demanding a finite-energy configuration \cite{francia2018TwoFormAsymptoticSymmetries,heissenberg2019TopicsAsymptoticSymmetries}.

    We can write the Lagrangian displacement in terms of \(\mathcal{H}\). Notice 
    \begin{equation}
        \pdv{\tensor{\xi}{^i}}{t} = \tensor{\var v}{^i},
    \end{equation}
    and thus 
    \begin{equation}
        \pdv{\tensor{\xi}{^i}}{t} = - \frac{1}{2} \tensor{\epsilon}{^i^j^k} \tensor{\mathcal{H}}{_u_j_k}.
    \end{equation}
    Using the chain rule we find
    \begin{equation}
        \pdv{\tensor{\xi}{^i}}{u} = - \pdv{\tensor{\xi}{^i}}{r} - \frac{1}{2} \tensor{\epsilon}{^i^j^k} \tensor{\mathcal{H}}{_u_j_k}.
    \end{equation}
    If we consider the falloffs for \(\tensor{\mathcal{H}}{_u_j_k}\) and for \(\tensor{\epsilon}{^i^j^k}\), and that the radial derivative of \(\tensor{\xi}{^i}\) must fall off faster than the \(u\)-derivative, we find
    \begin{equation}
        \pdv{\tensor{\xi}{^i}}{u} = - \frac{1}{2} \tensor{\epsilon}{^i^A^B} \tensor{\mathcal{H}}{_u_A_B} + o\qty(\frac{1}{r}).
    \end{equation}
    Notice \(\tensor{\epsilon}{^i^A^B} \sim 1/r^2\). Integrating, we obtain
    \begin{equation}\label{eq: memory-two-form-lagrangian-displacement}
        \Delta \tensor{\xi}{^i} = - \frac{1}{2}\tensor{\epsilon}{^i^A^B} \int_{-\infty}^{+\infty} \tensor{\mathcal{H}}{_u_A_B} \dd{u} + o\qty(\frac{1}{r}).
    \end{equation}
    Notice that using the expression for \(\tensor{H}{_u_A_B}\) in terms of \(\phi\) we recover
    \begin{equation}
        \Delta \tensor{\xi}{^i} = - \tensor{\delta}{^i^r} \Delta\phi + o\qty(\frac{1}{r}).
    \end{equation}
    This result matches Eq. \eqref{eq: Delta-xi-from-phi} at leading order, which was obtained in terms of \(\phi\) and using standard hydrodynamics techniques. The possible differences at subleading orders are due to our general Ansatz in Eq. \eqref{eq: falloffs-H}, which does not assume \(\mathcal{H}\) to be expanded in a power series in \(1/r\).

    Equation \eqref{eq: memory-two-form-lagrangian-displacement} plays a key role. Notice it critically depends on the interpretation of the two-form field as a dual description of linear acoustics. It is not a prediction from ``ab initio'' two-form electrodynamics. In fact, the natural coupling of a two-form field would be to a string, not to a point particle, which is how the Kalb--Ramond field finds its way into string theory \cite{polchinski1998IntroductionBosonicString,basile2025LecturesQuantumGravity,maccaferri2025IntroductionStringTheory}. Furthermore, a string memory effect has already been obtained in connection with the two-form theory \cite{afshar2019StringMemoryEffect}. Very importantly, however, Eq. \eqref{eq: memory-two-form-lagrangian-displacement} states we are considering how the two-form field is affecting the dynamics of the fluid particles, which ensures we are still analyzing an acoustic problem.

    Under the light of the electromagnetic analysis, Eq. \eqref{eq: memory-two-form-lagrangian-displacement} invites us to define a memory tensor through
    \begin{equation}\label{eq: memory-tensor-sound}
        \tensor{\Delta}{_A_B} = \int_{-\infty}^{+\infty} \tensor{H}{_u_A_B} \dd{u},
    \end{equation}
    where \(\tensor{H}{_u_A_B}\) is the leading-order contribution to \(\tensor{\mathcal{H}}{_u_A_B}\). Notice other components of \(\mathcal{H}\) do not enter the memory effect at leading order, and thus \(\tensor{\Delta}{_u_\mu} = \tensor{\Delta}{_r_\mu} = 0\). As in the electromagnetic case, the memory tensor cannot depend on \(u\) or \(r\), and it is a differential form on the sphere. 

    There is, however, an important difference at this point. For the acoustic two-form theory, the memory tensor is a two-form. Since it is a form on the sphere, it is automatically closed, because all three-forms on the sphere vanish. This, however, does not imply \(\tensor{\Delta}{_A_B}\) to be exact. The sphere has exactly one two-form which is closed, but not exact---its volume form. In general, \(\tensor{\Delta}{_A_B}\) will thus be given by
    \begin{equation}\label{eq: sound-memory-hodge-decomposed}
        \tensor{\Delta}{_A_B} = 2 \tensor{\mathcal{D}}{_[_A}\tensor{C}{_B_]} + C \tensor{\epsilon}{_A_B},
    \end{equation}
    where \(\tensor{C}{_A}\) is a one-form on the sphere, \(C\) is a constant real number, and \(\tensor{\epsilon}{_A_B}\) is the volume form on the sphere.

    Let us write the memory tensor in terms of the two-form potential. We have
    \begin{equation}
        \mathcal{H} = \dd\mathcal{B}.
    \end{equation}
    We assume the falloffs on \(\mathcal{B}\) to be \cite{manzoni2026HigherorderpformAsymptotic,campiglia2019ScalarAsymptoticCharges}
    \begin{subequations}\label{eq: falloffs-B-field}
        \begin{align}
            \tensor{\mathcal{B}}{_u_r} &= \frac{\tensor{B}{_u_r}}{r^2} + \tensor{\partial}{_u}\tensor{\mathcal{A}}{_r} - \tensor{\partial}{_r}\tensor{\mathcal{A}}{_u} + o\qty(\frac{1}{r^2}), \\
            \tensor{\mathcal{B}}{_r_A} &= \frac{\tensor{B}{_r_A}}{r} + \tensor{\partial}{_r}\tensor{\mathcal{A}}{_A} - \tensor{\mathcal{D}}{_A}\tensor{\mathcal{A}}{_r} + o\qty(\frac{1}{r}), \\
            \tensor{\mathcal{B}}{_u_A} &= \tensor{B}{_u_A} + \tensor{\partial}{_u}\tensor{\mathcal{A}}{_A} - \tensor{\mathcal{D}}{_A}\tensor{\mathcal{A}}{_u} + o\qty(1), \\
            \tensor{\mathcal{B}}{_A_B} &= r \tensor{B}{_A_B} + \tensor{\mathcal{D}}{_A}\tensor{\mathcal{A}}{_B} - \tensor{\mathcal{D}}{_B}\tensor{\mathcal{A}}{_A} + o\qty(r),
        \end{align}
    \end{subequations}
    where \(\mathcal{A}\) is a one-form with the expansion \cite{manzoni2026HigherorderpformAsymptotic}
    \begin{subequations}\label{eq: polyhomogeneous-A}
        \begin{align}
            \tensor{\mathcal{A}}{_u} &\sim \sum_{k=0} \frac{\tensor*{A}{^{(k)}_u}}{r^k} + \sum_{k=1} \frac{\tensor*{\hat{A}}{^{(k)}_u} \log{r}}{r^k}, \\
            \tensor{\mathcal{A}}{_r} &\sim \sum_{k=0} \frac{\tensor*{A}{^{(k)}_r}}{r^k} + \sum_{k=1} \frac{\tensor*{\hat{A}}{^{(k)}_r} \log{r}}{r^k}, \\
            \tensor{\mathcal{A}}{_A} &\sim \sum_{k=0} \frac{\tensor*{A}{^{(k)}_A}}{r^{k-1}} + \sum_{k=1} \frac{\tensor*{\hat{A}}{^{(k)}_A} \log{r}}{r^{k-1}}.
        \end{align}
    \end{subequations}
    This is analogous to the expansions we considered in Eqs. \eqref{eq: falloffs-A-field} and \eqref{eq: polyhomogeneous-lambda} for electrodynamics.
    
    We first find
    \begin{equation}
        \tensor{\mathcal{H}}{_u_A_B} = \tensor{\partial}{_u}\tensor{\mathcal{B}}{_A_B} + 2 \tensor{\mathcal{D}}{_[_A}\tensor{\mathcal{B}}{_B_]_u},
    \end{equation}
    and, at leading order,
    \begin{equation}
        \tensor*{H}{_u_A_B} = \tensor{\partial}{_u}\tensor*{B}{_A_B}.
    \end{equation}
    Integrating shows that
    \begin{equation}\label{eq: memory-tensor-shift-BAB}
        \tensor{\Delta}{_A_B} = \Delta \tensor{B}{_A_B}.
    \end{equation}
    It then follows from the memory tensor being closed that  
    \begin{equation}
        \Delta\tensor*{B}{_A_B} = 2 \tensor{\mathcal{D}}{_[_A}\tensor{C}{_B_]} + C \tensor{\epsilon}{_A_B}.
    \end{equation}
    Therefore, the difference between \(\tensor{B}{_A_B}\) at late and early times is a (large) gauge transformation parameterized by forms on the sphere, in addition to a global shift by a closed, but nonexact, form. Notice that \(\tensor{B}{_A_B}\) is multiplied by \(1/r\) in the expression for \(\mathcal{B}\). Hence, a closed, but nonexact, form in \(\tensor{B}{_A_B}\) is not a topological term in the full theory. In particular the constant \(C\) occurs in the expressions for \(\mathcal{H}\), as we shall now see. A topological contribution such as what happens in the Aharonov--Bohm effect or such as the configuration alluded to in Sec. \ref{subsec: gauge-potential} would be unrelated to the constant \(C\) above.

    The exact contributions are precisely what we would expect given our experience with electrodynamics. Let us investigate the nonexact term in more detail. We see that 
    \begin{equation}
        \Delta\tensor{B}{_A_B} = C \tensor{\epsilon}{_A_B} + \text{exact}.
    \end{equation}
    If we integrate both sides over the sphere, the exact piece will drop out---Stokes' theorem forces its integral to vanish. We find, however, that 
    \begin{equation}
        \Delta \oint_{\mathbb{S}^2} \tensor{B}{_A_B} = 4\pi C.
    \end{equation}
    This can also be written in terms of \(\mathcal{H}\). We find 
    \begin{equation}
        \int_{-\infty}^{+\infty} \oint_{\mathbb{S}^2} \tensor*{H}{_u_A_B} \dd{u} = 4\pi C.
    \end{equation}
    Use Eq. \eqref{eq: HuAB-from-phi} to rewrite the integral in terms of \(\phi\). At leading order, we find
    \begin{subequations}
        \begin{align}
            \int_{-\infty}^{+\infty} \oint_{\mathbb{S}^2} \tensor{\epsilon}{_A_B} \tensor{\partial}{_u}\phi^{(1)} \dd{u} &= 4\pi C, \\
            \oint_{\mathbb{S}^2} \Delta\phi^{(1)} \tensor{\epsilon}{_A_B} &= 4\pi C,
        \end{align}
    \end{subequations}
    where \(\phi^{(1)}\) is the \(1/r\) coefficient for \(\phi\) in a large-\(r\) expansion. Therefore, \(C\) measures the change in \(\phi\)'s monopole contribution. Notably, since \(\Delta \phi^{(1)}\) is independent of both \(u\) and \(r\), we find 
    \begin{equation}
        \Delta\phi^{(1)} = C + \text{coexact}.
    \end{equation}
    Notice coexact functions correspond to functions whose average over the sphere vanishes.

    At last, let us compute the exact part of the memory. The equations of motion imply
    \begin{equation}
        \tensor{\partial}{_u}\tensor{\mathcal{H}}{_B_u_r} - \tensor{\partial}{_r}\tensor{\mathcal{H}}{_B_u_r} - \frac{\tensor{\mathcal{D}}{^A}\tensor{\mathcal{H}}{_A_B_u}}{r^2} = 0.
    \end{equation}
    At leading order this means
    \begin{equation}
        \tensor{\partial}{_u}\tensor{H}{_B_u_r} = \tensor{\mathcal{D}}{^A}\tensor{H}{_A_B_u},
    \end{equation}
    and, upon integrating,
    \begin{equation}\label{eq: memory-tensor-sound-from-coulombic-field}
        \Delta\tensor{H}{_B_u_r} = \tensor{\mathcal{D}}{^A}\tensor{\Delta}{_A_B}.
    \end{equation}
    Since \(\tensor{\Delta}{_A_B}\) is closed, we know we can write this expression as 
    \begin{subequations}
        \begin{align}
            \Delta\tensor{H}{_A_u_r} &= 2\tensor{\mathcal{D}}{^B}\tensor{\mathcal{D}}{_[_B}\tensor{C}{_A_]}, \\
            &= (\mathcal{D}^2 - 1)\tensor{C}{_A} - \tensor{\mathcal{D}}{_A}\tensor{\mathcal{D}}{^B}\tensor{C}{_B}.
        \end{align}
    \end{subequations}
    Since we only consider the exterior derivative of \(\tensor{C}{_A}\), we have to take into account the gauge-for-gauge symmetry in \(\tensor{C}{_A}\). Fixing Lorenz gauge with \(\tensor{\mathcal{D}}{^A}\tensor{C}{_A} = 0\), we find 
    \begin{equation}
        \Delta\tensor{H}{_A_u_r} = (\mathcal{D}^2 - 1)\tensor{C}{_A}.
    \end{equation}
    This expression can now be inverted, and thus \(\tensor{C}{_A}\) is uniquely defined in terms of the change in the ``Coulombic field,'' \(\Delta\tensor{H}{_A_u_r}\). Notice the solution is only unique once we have fixed \(\tensor{\mathcal{D}}{^A}\tensor{C}{_A} = 0\).

\section{Asymptotic Symmetries}\label{sec: asymptotic-symmetries}
    We consider the two-form theory in Lorenz gauge. Recall, as discussed in Sec. \ref{subsec: gauge-potential}, of the presence of reducible gauge symmetries. 

    We know we can write the field-strenght \(\mathcal{H}\) in terms of a two-form potential
    \begin{equation}
        \mathcal{H} = \dd\mathcal{B}.
    \end{equation}
    In Lorenz gauge, this two-form potential is required to obey
    \begin{equation}
        \dd\hodge\mathcal{B} = 0.
    \end{equation}
    The residual gauge symmetries are the one-forms obeying 
    \begin{equation}
        \dd\hodge\dd\mathcal{A} = 0,
    \end{equation}
    which we take to be already gauge-fixed in Lorenz gauge
    \begin{equation}
        \dd\hodge\mathcal{A} = 0.
    \end{equation}
    Because the original theory has reducible gauge symmetries, these residual gauge symmetries also have residual gauge-for-gauge symmetry, which is parameterized by scalar functions obeying 
    \begin{equation}
        \dd\hodge\dd\lambda = 0.
    \end{equation}

    As in the electromagnetic case, there are subtleties requiring the use of logarithms in the large-\(r\) expansions for the gauge parameters---see Refs. \cite{manzoni2026HigherorderpformAsymptotic,francia2024AsymptoticChargespforms,satishchandran2019AsymptoticBehaviorMassless}. This is why we will use the expansion in Eq. \eqref{eq: polyhomogeneous-A}. 

    We want to find the most general residual gauge transformations that survive at infinity. These are the asymptotic symmetries. Since the gauge parameter \(\mathcal{A}\) obeys \(\dd\hodge\dd\mathcal{A} = 0\) in Lorenz gauge, we are essentially solving the Maxwell equations in vacuum.

    For the moment, let us focus on the bulk and ignore the ``survive at infinity'' part. We first write the Lorenz gauge condition. Define a scalar field \(\mathcal{L}\) by
    \begin{equation}
        \mathcal{L} = \dd\hodge\mathcal{A}.
    \end{equation}
    Then ``Lorenz gauge'' means \(\mathcal{L} = 0\). In Bondi coordinates, we have
    \begin{equation}\label{eq: definition-mathcal-L}
        \mathcal{L} = \tensor{\partial}{_u}\tensor{\mathcal{A}}{_r} + \frac{1}{r^2}\pdv{r}\qty(r^2 \tensor{\mathcal{A}}{_u}) - \frac{1}{r^2}\pdv{r}\qty(r^2 \tensor{\mathcal{A}}{_r}) - \frac{\tensor{\mathcal{D}}{^B}\tensor{\mathcal{A}}{_B}}{r^2} = 0.
    \end{equation}
    Meanwhile, the wave equation \(\dd\hodge\dd\mathcal{A}=0\) is given by
    \begin{widetext}
    \begin{subequations}
        \begin{gather}
            \frac{2}{r}\pdv{r}\qty(r\tensor{\partial}{_u}\tensor{\mathcal{A}}{_u}) - \frac{1}{r^2}\pdv{r}\qty(r^2\tensor{\partial}{_r}\tensor{\mathcal{A}}{_u}) -  \frac{\mathcal{D}^2\tensor{\mathcal{A}}{_u}}{r^2} - \tensor{\partial}{_u}\mathcal{L} = 0, \\
            \frac{2}{r}\pdv{r}\qty(r\tensor{\partial}{_u}\tensor{\mathcal{A}}{_r}) - \frac{1}{r^2}\pdv{r}\qty(r^2\tensor{\partial}{_r}\tensor{\mathcal{A}}{_r}) -  \frac{(\mathcal{D}^2-2)\tensor{\mathcal{A}}{_r}}{r^2} - \frac{2\tensor{\mathcal{A}}{_u}}{r^2} +\frac{2 \tensor{\mathcal{D}}{^B}\tensor{\mathcal{A}}{_B}}{r^3} - \tensor{\partial}{_r}\mathcal{L} = 0, \label{subeq: wave-Ar} \\
            2 \tensor{\partial}{_u}\tensor{\partial}{_r}\tensor{\mathcal{A}}{_A} - \tensor{\partial}{_r}\tensor{\partial}{_r}\tensor{\mathcal{A}}{_A} -  \frac{(\mathcal{D}^2-1)\tensor{\mathcal{A}}{_A}}{r^2} + \tensor{\mathcal{D}}{_A}\qty(\frac{2 \tensor{\mathcal{A}}{_u}}{r} - \frac{2 \tensor{\mathcal{A}}{_r}}{r}) - \tensor{\mathcal{D}}{_A}\mathcal{L} = 0.
        \end{gather}
    \end{subequations}

    At this point, it is tempting to use Eq. \eqref{eq: polyhomogeneous-A} to obtain order-by-order equations for the gauge parameter. For example, \(\mathcal{L}\) would be written as 
    \begin{equation}
        \mathcal{L} \sim \sum_{k=0} \frac{L^{(k)}}{r^k} + \sum_{k=1} \frac{\hat{L}^{(k)} \log{r}}{r^k},
    \end{equation}
    and matching order-by-order with the expansion for \(\mathcal{A}\) in Eq. \eqref{eq: definition-mathcal-L} would give 
    \begin{subequations}\label{eq: expansion-L}
        \begin{gather}
            L^{(n)} = \tensor{\partial}{_u}\tensor*{A}{^{(n)}_r} - (n-3)\tensor*{A}{^{(n-1)}_u} + \tensor*{\hat{A}}{^{(n-1)}_u} + (n-3)\tensor*{A}{^{(n-1)}_r} - \tensor*{\hat{A}}{^{(n-1)}_r} - \tensor{\mathcal{D}}{^B}\tensor*{A}{^{(n-1)}_B} \\
            \intertext{and}
            \hat{L}^{(n)} = \tensor{\partial}{_u}\tensor*{\hat{A}}{^{(n)}_r} - (n-3)\tensor*{\hat{A}}{^{(n-1)}_u} + (n-3)\tensor*{\hat{A}}{^{(n-1)}_r} - \tensor{\mathcal{D}}{^B}\tensor*{\hat{A}}{^{(n-1)}_B}.
        \end{gather}
    \end{subequations}
    We can then try expanding all components of the wave equation, and solving order-by-order.

    While this is logical, it is not practical. Notice \(L^{(n)}\), for example, depends on both \(\tensor*{A}{_r^{(n)}}\) and \(\tensor*{A}{_r^{(n-1)}}\), and similarly with other components. Hence, \(L^{(n)}\) is mixing different orders (\(n\) and \(n-1\)) in the same object. This is very inconvenient to handle. Instead of handling order \(n-1\), and then moving on to handle order \(n\), we are forced to solve them all at once! To avoid this, notice that the only term in \(L^{(n)}\) that is of order \(n\), instead of \(n-1\) comes with a \(u\)-derivative. Since \(u\) and \(r\) have the same dimensions, dimensional analysis forces \(\tensor{\partial}{_u}\tensor*{A}{_r^{(n)}}\) to occur at the same order as \(\tensor*{A}{_r^{(n-1)}}\). We can, however, disentangle this behavior. 
    
    Adapting an analysis by \textcite{satishchandran2019AsymptoticBehaviorMassless}, we define a new scalar 
    \begin{equation}
        \mathcal{W} = \qty[\frac{2}{r}\pdv{r}\qty(r\tensor{\partial}{_u}\tensor{\mathcal{A}}{_r}) - \frac{1}{r^2}\pdv{r}\qty(r^2\tensor{\partial}{_r}\tensor{\mathcal{A}}{_r}) -  \frac{(\mathcal{D}^2-2)\tensor{\mathcal{A}}{_r}}{r^2} - \frac{2\tensor{\mathcal{A}}{_u}}{r^2} +\frac{2 \tensor{\mathcal{D}}{^B}\tensor{\mathcal{A}}{_B}}{r^3} - \tensor{\partial}{_r}\mathcal{L}] - \frac{1}{r^2}\pdv{r}\qty(r^2\mathcal{L}).
    \end{equation}
    The term in brackets is the \(r\)-component of the wave equation, Eq. \eqref{subeq: wave-Ar}. The advantage will be in that \(\mathcal{W} = 0\) is almost sufficient to impose Lorenz gauge (as we shall now see), but the order-by-order expansion of \(\mathcal{W}\) will not mix orders. 
    
    If we assume the wave equation to hold, then 
    \begin{equation}
        \mathcal{W} \approx - \frac{1}{r^2}\pdv{r}\qty(r^2\mathcal{L}),
    \end{equation}
    where ``\(\approx\)'' stands for ``equality on-shell.'' Therefore, once the wave equation is imposed, \(\mathcal{W}\) vanishes if, and only if, 
    \begin{equation}
        \pdv{r}\qty(r^2\mathcal{L}) = 0.
    \end{equation}
    In other words, once the wave equation is imposed, \(\mathcal{W}\) vanishes if, and only if, 
    \begin{equation}
        \mathcal{L} = \frac{L^{(2)}}{r^2}.
    \end{equation}
    Which means that \(\mathcal{W} \approx 0\) is almost the same as \(\mathcal{L} \approx 0\)! To ensure \(\mathcal{L} \approx 0\), we explicitly ask that \(L^{(2)}\) vanishes too. Hence, once the equations of motion are imposed, \(\mathcal{L}\) vanishes if, and only if, \(\mathcal{W}\) and \(L^{(2)}\) vanish. 
    
    The advantage of working with \(\mathcal{W}\) is that, after expanding the definition of \(\mathcal{L}\),
    \begin{equation}\label{eq: expression-mathcal-W}
        \mathcal{W} = \frac{1}{r^4}\pdv{r}\qty(r^4\tensor{\partial}{_r}\tensor{\mathcal{A}}{_r}) -  \frac{(\mathcal{D}^2-2)\tensor{\mathcal{A}}{_r}}{r^2} - \frac{2\tensor{\mathcal{A}}{_u}}{r^2} - \frac{2}{r^3}\pdv{r}\qty(r^3 \tensor{\partial}{_r}\tensor{\mathcal{A}}{_u}) + \frac{2 \tensor{\partial}{_r}\tensor{\mathcal{D}}{^B}\tensor{\mathcal{A}}{_B}}{r^2},
    \end{equation}
    which has no \(u\)-derivatives! As a consequence, \(W^{(n)}\) will not mix different orders.

    To check that is so, let us write
    \begin{equation}
        \mathcal{W} \sim \sum_{k=0} \frac{W^{(k)}}{r^k} + \sum_{k=1} \frac{\hat{W}^{(k)} \log{r}}{r^k}.
    \end{equation}
    Then Eqs. \eqref{eq: polyhomogeneous-A} and \eqref{eq: expression-mathcal-W} yield
    \begin{subequations}\label{eq: expansion-W}
        \begin{gather}
            W^{(n+2)} = - 2 (n-1) \tensor{\mathcal{D}}{^B}\tensor*{A}{^{(n)}_B} + 2 \tensor{\mathcal{D}}{^B}\tensor*{\hat{A}}{^{(n)}_B} - (2n-3) \tensor*{\hat{A}}{^{(n)}_r} + 4 (n-1)\tensor*{\hat{A}}{^{(n)}_u} - [\mathcal{D}^2 - (n-1)(n-2)]\tensor*{A}{^{(n)}_r} - 2 (n- 1)^2\tensor*{A}{^{(n)}_u} \\
            \intertext{and}
            \hat{W}^{(n+2)} = - 2 (n-1) \tensor{\mathcal{D}}{^B}\tensor*{\hat{A}}{^{(n)}_B} - [\mathcal{D}^2 - (n-1)(n-2)]\tensor*{\hat{A}}{^{(n)}_r} - 2 (n- 1)^2\tensor*{\hat{A}}{^{(n)}_u}.
        \end{gather}
    \end{subequations}
    As promised, no orders are mixed. To impose Lorenz gauge when solving the wave equation, we merely need to impose \(L^{(2)} = 0\) and \(W^{(n)} = 0\) for all \(n\).
    
    Next we expand the wave equation order by order. In this case, orders will indeed be mixed---the leading data at low \(n\) is used to determine the subleading data at higher \(n\). We have 
    \begin{subequations}\label{eq: wave-equation-A-expanded}
        \begin{gather}
            - 2 (n-1)\tensor{\partial}{_u}\tensor*{A}{^{(n)}_u} + 2 \tensor{\partial}{_u}\tensor*{\hat{A}}{^{(n)}_u} + (2n-3) \tensor*{\hat{A}}{^{(n-1)}_u} - [\mathcal{D}^2+(n-1)(n-2)]\tensor*{A}{^{(n-1)}_u} - \tensor{\partial}{_u}L^{(n+1)} = 0, \\
            - 2 (n-1)\tensor{\partial}{_u}\tensor*{\hat{A}}{^{(n)}_u} - [\mathcal{D}^2+(n-1)(n-2)]\tensor*{\hat{A}}{^{(n-1)}_u} - \tensor{\partial}{_u}\hat{L}^{(n+1)} = 0, \\
            -2(n-1)\tensor{\partial}{_u}\tensor*{A}{^{(n)}_r} + 2 \tensor{\partial}{_u}\tensor*{\hat{A}}{^{(n)}_r} + (2n-3) \tensor*{\hat{A}}{^{(n-1)}_r} - [\mathcal{D}^2+n(n-3)]\tensor*{A}{^{(n-1)}_r} - 2 \tensor*{A}{^{(n-1)}_u} + 2 \tensor{\mathcal{D}}{^B}\tensor*{A}{^{(n-1)}_B} + n L^{(n)} - \hat{L}^{(n)} = 0, \\
            -2(n-1)\tensor{\partial}{_u}\tensor*{\hat{A}}{^{(n)}_r} - [\mathcal{D}^2+n(n-3)]\tensor*{\hat{A}}{^{(n-1)}_r} - 2 \tensor*{\hat{A}}{^{(n-1)}_u} + 2 \tensor{\mathcal{D}}{^B}\tensor*{\hat{A}}{^{(n-1)}_B} + n \hat{L}^{(n)} = 0, \\
            - 2(n-1) \tensor{\partial}{_u}\tensor*{A}{^{(n)}_A} + 2\tensor{\partial}{_u}\tensor*{\hat{A}}{^{(n)}_A} + (2n-3)\tensor*{\hat{A}}{^{(n-1)}_A} - [\mathcal{D}^2 - 1 + (n-1)(n-2)]\tensor*{A}{^{(n-1)}_A} - 2 \tensor{\mathcal{D}}{_A}\tensor*{A}{^{(n-1)}_r} + 2 \tensor{\mathcal{D}}{_A}\tensor*{A}{^{(n-1)}_u} - \tensor{\mathcal{D}}{_A}L^{(n)} = 0, \\
            - 2(n-1) \tensor{\partial}{_u}\tensor*{\hat{A}}{^{(n)}_A} - [\mathcal{D}^2 - 1 + (n-1)(n-2)]\tensor*{\hat{A}}{^{(n-1)}_A} - 2 \tensor{\mathcal{D}}{_A}\tensor*{\hat{A}}{^{(n-1)}_r} + 2 \tensor{\mathcal{D}}{_A}\tensor*{\hat{A}}{^{(n-1)}_u} - \tensor{\mathcal{D}}{_A}\hat{L}^{(n)} = 0.
        \end{gather}
    \end{subequations}
    \end{widetext}
    These expressions match Eqs. (28)--(32) in Ref. \cite{satishchandran2019AsymptoticBehaviorMassless}, up to our addition of an extra series with \(\log{r}/r^k\) terms. 

    Let us consider the implications at the first few orders. To impose Lorenz gauge, we will need to impose \(L^{(2)} = 0\). This condition is 
    \begin{equation}\label{eq: lorenz-L-2}
        \tensor{\partial}{_u}\tensor*{A}{^{(2)}_r} +\tensor*{A}{^{(1)}_u} + \tensor*{\hat{A}}{^{(1)}_u} -\tensor*{A}{^{(1)}_r} - \tensor*{\hat{A}}{^{(1)}_r} - \tensor{\mathcal{D}}{^B}\tensor*{A}{^{(1)}_B} = 0.
    \end{equation}
    Apart from this condition, we can solve the system by working order-by-order with the wave equations and \(\mathcal{W}\).
    
    Let us write the equations involving the leading-order data for \(\mathcal{A}\). The gauge condition \(W^{(2)} = 0\) imposes
    \begin{equation}\label{eq: constraint-leading-order-A}
        2 \tensor{\mathcal{D}}{^B}\tensor*{A}{^{(0)}_B} - [\mathcal{D}^2 - 2]\tensor*{A}{^{(0)}_r} - 2\tensor*{A}{^{(0)}_u} = 0.
    \end{equation}
    If we set \(n=0\) on Eq. \eqref{eq: wave-equation-A-expanded} and impose the Lorenz gauge condition (the latter being equivalent to \(L^{(2)} = 0\) and \(\mathcal{W} = 0\)), we find
    \begin{subequations}\label{eq: leading-A-large-gauge-time-independent}
        \begin{gather}
            \tensor{\partial}{_u}\tensor*{A}{^{(0)}_u} = 0, \\
            \tensor{\partial}{_u}\tensor*{A}{^{(0)}_r} = 0, \\
            \tensor{\partial}{_u}\tensor*{A}{^{(0)}_A} = 0.
        \end{gather}
    \end{subequations}
    We thus see that the leading components of \(\mathcal{A}\) are required to be time-independent. 
    
    Can higher-order terms further constrain these initial data? For \(n=1\), the gauge condition \(\mathcal{W} = 0\) implies 
    \begin{subequations}
        \begin{gather}
            W^{(3)} = 2 \tensor{\mathcal{D}}{^B}\tensor*{\hat{A}}{^{(1)}_B} + \tensor*{\hat{A}}{^{(1)}_r} - \mathcal{D}^2\tensor*{A}{^{(1)}_r} = 0 \\
            \intertext{and}
            \hat{W}^{(3)} = - \mathcal{D}^2\tensor*{\hat{A}}{^{(1)}_r} = 0.
        \end{gather}
    \end{subequations}
    The equation \(\mathcal{D}^2\tensor*{\hat{A}}{^{(1)}_r} = 0\) means \(\tensor*{\hat{A}}{^{(1)}_r}\) is constant on the sphere. However, the imposition \(W^{(3)} = 0\) forces \(\tensor*{\hat{A}}{^{(1)}_r}\) to have vanishing average on the sphere. Thus, \(\tensor*{\hat{A}}{^{(1)}_r} = 0\). In the wave equations we have, at \(n=1\) and Lorenz gauge,
    \begin{subequations}\label{eq: need-for-logarithms-for-A}
        \begin{gather}
            2 \tensor{\partial}{_u}\tensor*{\hat{A}}{^{(1)}_u} - \mathcal{D}^2\tensor*{A}{^{(0)}_u} = 0, \\
            2 \tensor{\partial}{_u}\tensor*{\hat{A}}{^{(1)}_r} - [\mathcal{D}^2-2]\tensor*{A}{^{(0)}_r} - 2 \tensor*{A}{^{(0)}_u} + 2 \tensor{\mathcal{D}}{^B}\tensor*{A}{^{(0)}_B} = 0, \\
            2 \tensor{\partial}{_u}\tensor*{\hat{A}}{^{(1)}_A} - [\mathcal{D}^2 - 1]\tensor*{A}{^{(0)}_A} - 2 \tensor{\mathcal{D}}{_A}\tensor*{A}{^{(0)}_r} + 2 \tensor{\mathcal{D}}{_A}\tensor*{A}{^{(0)}_u} = 0.
        \end{gather}
    \end{subequations}
    We already know \(\tensor*{\hat{A}}{^{(1)}_r} = 0\), and indeed Eq. \eqref{eq: constraint-leading-order-A} means the expression for \(\tensor{\partial}{_u}\tensor*{\hat{A}}{^{(1)}_r}\) vanishes, as it should. The remaining equations are differential equations setting the time-evolution of \(\tensor*{\hat{A}}{^{(1)}_u}\) and \(\tensor*{\hat{A}}{^{(1)}_A}\) in terms of previously known data. We see that no new conditions for the leading data on \(\mathcal{A}\) appear. This would not be the case had we not added the logarithmic series in Eq. \eqref{eq: polyhomogeneous-A}, because then Eq. \eqref{eq: need-for-logarithms-for-A} would constrain \(\tensor*{A}{^{(0)}_u}\), \(\tensor*{A}{^{(0)}_r}\), and \(\tensor*{A}{^{(0)}_A}\) even more. 
    
    As we progress for larger values of \(n\), the subleading data is determined by the leading data, as in the electromagnetic case.

    We still have residual gauge-to-gauge freedom in \(\mathcal{A}\). If we shift 
    \begin{equation}
        \mathcal{A} \to \mathcal{A} + \dd\lambda,
    \end{equation}
    with \(\dd\hodge\dd\lambda = 0\), we still preserve Lorenz gauge. Pick 
    \begin{equation}
        \lambda \sim r \lambda^{(-1)} + \log{r} \hat{\lambda}^{(0)} + \lambda^{(0)} + o(1),
    \end{equation}
    which is more general than Eq. \eqref{eq: polyhomogeneous-lambda} because we are admitting larger falloffs for \(\mathcal{A}\) in comparison to what we did in electrodynamics. Then, the first few orders of the wave equation for \(\lambda\) yield
    \begin{subequations}
        \begin{align}
            \tensor{\partial}{_u}\lambda^{(-1)} = 0, \\
            \tensor{\partial}{_u}\lambda^{(0)} = \frac{1}{2}\qty[\mathcal{D}^2 + 6]\lambda^{(-1)}, \\
            \tensor{\partial}{_u}\hat{\lambda}^{(0)} = 0,
        \end{align}
    \end{subequations}
    and so on. In practice, we can obtain residual gauge-for-gauge transformations with the behavior
    \begin{equation}
        \lambda = r\varepsilon(\tensor{x}{^A}) + \log{r} \hat{\varepsilon}(\tensor{x}{^A}) + \order{1}.
    \end{equation}
    By setting \(\varepsilon = - \tensor*{A}{^{(0)}_r}\), we can eliminate \(\tensor*{A}{^{(0)}_r}\) from the leading order data. Equation \eqref{eq: constraint-leading-order-A} then determines \(\tensor*{A}{^{(0)}_u}\) from \(\tensor*{A}{^{(0)}_A}\), which we are free to specify up to an exact form (which is the gauge transformation \(\lambda\)).

    Hence, we notice the existence of nontrivial residual gauge transformations that survive at infinity. These are given by 
    \begin{equation}\label{eq: large-gauge-A-at-large-r}
        \tensor{\mathcal{A}}{_\mu}\dd{\tensor{x}{^\mu}} = r \tensor*{A}{^{(0)}_A} \dd{\tensor{x}{^A}} + o(r).
    \end{equation}

    In terms of \(\mathcal{B}\), we see these large gauge transformations act on \(\tensor{B}{_A_B}\) according to
    \begin{equation}
        \tensor{B}{_A_B} \to \tensor{B}{_A_B} + 2\tensor{\mathcal{D}}{_[_A}\tensor*{A}{^{(0)}_B_]}.
    \end{equation}
    All other components of \(\mathcal{B}\) are shifted in orders smaller than \(r\).

    Recall that, by construction, the acoustic memory tensor \eqref{eq: memory-tensor-sound} is \(r\)-independent, \(u\)-independent, and has \(\tensor{\Delta}{_u_\mu} = \tensor{\Delta}{_r_\mu} = 0\). Then, using Eqs. \eqref{eq: sound-memory-hodge-decomposed} and \eqref{eq: memory-tensor-shift-BAB}, we finally notice that the memory tensor can be written as 
    \begin{equation}\label{eq: memory-tensor-large-gauge-kalb-ramond}
        r\tensor{\Delta}{_\mu_\nu} = \tensor{\partial}{_\mu}\tensor{\mathcal{A}}{_\nu} - \tensor{\partial}{_\nu}\tensor{\mathcal{A}}{_\mu} + C \tensor{\epsilon}{_\mu_\nu} + o(r),
    \end{equation}
    where \(\tensor{\mathcal{A}}{_\mu}\) is a large gauge transformation for the two-form description of acoustics, \(C\) is a constant, and \(\tensor{\epsilon}{_\mu_\nu}\) is the volume form on the unit sphere. Hence, the memory tensor can be written as a large gauge transformation, except in the particular case of monopole-induced memory. 

\section{Soft Theorem}\label{sec: soft-theorem}
    The asymptotic symmetries we obtained are the same ones analyzed by \textcite{campiglia2019ScalarAsymptoticCharges,francia2018TwoFormAsymptoticSymmetries}. Their goal was to understand, in terms of a two-form, the asymptotic symmetries that would originate a scalar soft theorem \cite{campiglia2018CanScalarsHave}. Hence, the connection between the asymptotic symmetries discussed above and a scalar soft theorem is already known. In particular, \textcite{campiglia2019ScalarAsymptoticCharges} highlight how the \(C\)-term in our memory tensor receives a different treatment in the two-form description. It is associated to a symmetry that cannot be written as a large gauge transformation in general, and has no interpretation in the two-form theory. Indeed, our interpretation above of this term was as a permanent shift in the monopole of the scalar field \(\phi\). 

    Our task is then to find a link between this soft theorem and the memory effect. A key remark is that soft theorems are typically connected to memory effects via a Braginsky--Thorne--like formula \cite{braginsky1987GravitationalwaveBurstsMemory,strominger2016GravitationalMemoryBMS}. See, for example, Sec. 6.6 in Ref. \cite{aguiaralves2026LecturesBondiMetzner}. This is very natural for theories coupled to charged particles, such as gravitation and electrodynamics. For sound, however, the sources we consider are merely ``image sources,'' not physical particles in a quantum or classical sense. 

    If we overlook this fact, we can consider a set of pointlike sources and study the memory effect associated to them. For one pointlike source following a trajectory \(\vb{z}(t)\), we pick
    \begin{equation}
        q(t,\vb{r}) = \frac{Q}{\sqrt{1 - v^2/c^2}} \delta^{(3)}(\vb{r} - \vb{z}(t)),
    \end{equation}
    where \(v\) is the norm of the velocity of the particle and \(Q\) is a constant. Notice we effectively introduced a ``sonic'' Lorentz factor in the expression for the density of a pointlike charge. In relativity, this is the combination that would be Lorentz invariant for a scalar charge. It originates from 
    \begin{subequations}
        \begin{align}
            q(x) &= Q \int \delta^{(4)}(x - z(\tau)) \dd{\tau}, \\
            &= Q \int \delta(t - \tensor{z}{^0}(\tau)) \delta^{(3)}(\vb{r} - \vb{z}(\tau)) \dd{\tau}, \\
            &= Q \dv{\tau}{t} \delta^{(3)}(\vb{r} - \vb{z}(t)).
        \end{align}
    \end{subequations}
    
    If we choose to consider such a source, then the wave equation becomes 
    \begin{equation}
        -\frac{1}{c^2}\pdv[2]{\phi}{t} + \laplacian\phi = \frac{Q}{\sqrt{1 - v^2/c^2}} \delta^{(3)}(\vb{r} - \vb{z}(t)),
    \end{equation}
    and the retarded solution is given by 
    \begin{equation}
        \phi(t,\vb{r}) = - \frac{Q c}{4 \pi \sqrt{c^2 - v^2}} \frac{c}{c - \vu{n} \vdot \vb{v}} \frac{1}{\norm{\vb{r} - \vb{z}(t_{\text{ret}})}},
    \end{equation}
    where 
    \begin{equation}
        \vu{n} = \frac{\vb{r} - \vb{z}(t_{\text{ret}})}{\norm{\vb{r} - \vb{z}(t_{\text{ret}})}},
    \end{equation}
    \(t_{\text{ret}}\) is defined by 
    \begin{equation}
        c(t - t_{\text{ret}}) = \norm{\vb{r} - \vb{z}(t_{\text{ret}})},
    \end{equation}
    and the velocities are always evaluated at retarded time. This construction may seem arbitrary, but remember the monopole \(Q\) is not a fundamental property of the sources. The sources are not particles in any sense. Rather, they are a convenient way of phrasing boundary conditions through the method of images. In introducing a Lorentz-like factor, we are effectively redefining the meaning of ``monopole'' in a way that will be convenient later on. For an inertial particle, this just means multiplying \(Q\) by a constant.  

    If we consider a scattering process with a number of incoming sources, and a number of outgoing sources, then the memory in \(\phi\) is 
    \begin{multline}\label{eq: braginsky-thorne}
        \Delta\phi(\vb{r}) = - \frac{c^2}{4\pi r} \left[\sum_{\substack{i\\\text{incoming}}} \frac{Q_i}{\sqrt{c^2 - v_i^2}(c - \vu{r} \cdot \vb{v}_i)}\right. \\ \left. {} - \sum_{\substack{j\\\text{outgoing}}} \frac{Q_j}{\sqrt{c^2 - v_j^2}(c - \vu{r} \cdot \vb{v}_j)}\right].
    \end{multline}
    Because the sources are not fundamental, this expression is not as general and far-reaching as the original Braginsky--Thorne formula \cite{braginsky1987GravitationalwaveBurstsMemory}. Nevertheless, it still bears physical meaning on a finely-tuned experiment. For instance, picture a process in which several gas bubbles are thrown in. Each of them expanding or contracting so it has constant \(Q_i\). The bubbles interact in some complicated way, and at the end a number of bubbles moves away from the interaction region (also with asymptotically constant \(Q_j\)). Then Eq. \eqref{eq: braginsky-thorne} would describe the acoustic memory in this finely-tuned experiment. As a simpler particular case, Eq. \eqref{eq: braginsky-thorne} can describe the memory due to a pulsating sphere---such as the one described at the beginning of the paper.

    The denominators in Eq. \eqref{eq: braginsky-thorne} can be rewritten as 
    \begin{equation}\label{eq: denominator-braginsky-thorne}
        \sqrt{c^2 - v^2}(c - \vu{r} \cdot \vb{v}) = -\tensor{u}{^\mu}\tensor{k}{_\mu},
    \end{equation}
    where \(\tensor{u}{^\mu}\) is akin to a four-velocity, with components
    \begin{equation}
        \tensor{u}{^\mu} = \sqrt{1 - \frac{v^2}{c^2}} (c, \vb{v}),
    \end{equation}
    while \(\tensor{k}{^\mu}\) is a null vector pointing in the direction of \(\vu{r}\),
    \begin{equation}
        \tensor{k}{^\mu} = \qty(1, \vu{r}).
    \end{equation}
    Above, we considered the coordinate basis to be \((ct,x,y,z)\), for simplicity in writing the components. 

    Given Eq. \eqref{eq: denominator-braginsky-thorne}, one can see that the kinematical combinations that occur in the soft theorems considered by \textcite{campiglia2018CanScalarsHave} are precisely of the form of Eq. \eqref{eq: braginsky-thorne}. Hence, with a suitable choice of coupling constants, the acoustic Braginsky--Thorne formula can be reproduced from a soft theorem in a quantum field theory. We omit the detailed calculations, since they are completely analogous to the standard results for other field theories---see, e.g., Sec. 6.6 in Ref. \cite{aguiaralves2026LecturesBondiMetzner}.

\section{Discussion}\label{sec: discussion}
    We discussed an infrared triangle for linear perturbations in inviscid, irrotational, barotropic fluid flow. We have shown how acoustics can present a memory effect, and how a dual formulation in terms of a two-form allows us to understand this memory effect in terms of asymptotic symmetries. These asymptotic symmetries, in turn, can be associated to soft scalar theorems, which can also be used to reproduce the memory effect. Two key consequences of this analysis are the following:
    \begin{enumerate}
        \item Asymptotic symmetries and the infrared triangle are not restricted to the realm of high-energy physics, and can also be realized in condensed matter systems \cite{perez2024FractonInfraredTriangle}. Notably, this is true even for systems as simple as linear sound. 
        \item In addition to the experimental prospects to detect gravitational wave memory \cite{grant2023OutlookDetectingGravitationalwave} and electromagnetic memory \cite{bieri2024ExperimentMeasureElectromagnetic}, it is also possible to expect measurements of acoustic memory in a laboratory. While one possible route is the effect identified by \textcite{datta2022InherentNonlinearityFluid}, we notice the memory effect described above is qualitatively different. Furthermore, the observation of such effects would also admit an interpretation in terms of asymptotic symmetries.
    \end{enumerate}

    Several comments are in order at this point. First and foremost, we discuss why a memory effect for sound is expected. The key remark is that Eqs. \eqref{eq: retarded-solution-phi}, \eqref{eq: large-r-integral-phi}, and \eqref{eq: general-solution-phi} are not specific to sound. They are all derived considering a wave equation, using the retarded Green's function, and finally performing a large-\(r\) expansion. One immediately sees from Eq. \eqref{eq: general-solution-phi} that if \(Q\) asymptotes to different constants at early and late retarded times, then there will be a memory effect in the sense that \(\Delta \phi\) will be nonzero at order \(1/r\). This is not a property of any specific theory---be it general relativity, Yang--Mills, electrodynamics, acoustics, or else---but rather a very general property of the wave equation in four dimensions. There are no nonlinear effects involved and this memory shift does not depend on a probe in any way. It is the response of the field to a permanent change in the source. 

    Relativistic theories are often coupled to conserved sources. For example, electric charge is conserved, and so is energy-momentum. This leads to particularly well-behaved memory effects, which can naturally arise in scattering. Since electric charges or masses cannot be created nor destroyed, simply changing their kinematic configurations can be enough to create memory. For instance, the original gravitational-wave displacement memory effect was conceived by imagining a cluster of stars orbiting the center of a galaxy \cite{zeldovich1974RadiationGravitationalWaves,braginsky1985KinematicResonanceMemory,braginsky1987GravitationalwaveBurstsMemory}. Acoustics, on the other hand, is not coupled to a conserved source. A sphere with constant radius, for instance, has a vanishing monopole moment, and does not produce any kind of sound. This is a sense in which memory in effective theories can be unusual, because physical realizations may need to handle the sources with more care than in their fundamental counterparts.

    Once it is clear that a memory effect for sound can be obtained---which, as stated, is a direct consequence of the wave equation---one can inquire about connections to the other corners of an infrared triangle. Within the approximations we worked on, acoustic perturbations are mathematically identical to a massless Klein--Gordon field. This allowed us to explore a duality between the Klein--Gordon field and a two-form theory, which had already been investigated before in the context of asymptotic symmetries \cite{campiglia2019ScalarAsymptoticCharges,francia2018TwoFormAsymptoticSymmetries}. Nevertheless, to our knowledge, no connections to a memory effect had been noticed yet. 

    Given the previous work by \textcite{campiglia2019ScalarAsymptoticCharges,francia2018TwoFormAsymptoticSymmetries}, it is expected that scalar fields admit asymptotic symmetries in some sense, and thus that this could be carried over to sound. Nevertheless, it is nontrivial a priori that these symmetries can still describe the long-distance behavior predicted by the Euler equations. This nontriviality is due to how we define ``memory.'' For a relativistic scalar field, the ``natural'' memory effect would be a momentum kick to a probe, as discussed by \textcite{tolish2014RetardedFieldsNull}. We, however, defined the ``acoustic memory effect'' to be a permanent shift in the Lagrangian displacement of a fluid particle. This is physically very different from the momentum kick. Yet, both effects can be tracked down to a permanent shift in the scalar field, which in turn could be expected to be explained in terms of dual asymptotic symmetries. 

    There is one notable exception to the asymptotic symmetry interpretation. In Sec. \ref{sec: memory-kalb-ramond} we highlighted how the memory tensor is not an exact form on the sphere, and in Sec. \ref{sec: asymptotic-symmetries} we only related the exact part of the memory tensor to asymptotic symmetries. The nonexact part was instead associated, in Sec. \ref{sec: memory-kalb-ramond}, to a permanent shift in the monopole term of \(\phi\). The same sort of difficulty was identified by \textcite{campiglia2019ScalarAsymptoticCharges}, albeit in that case they were concerned with interpreting the surface charges of a scalar theory in terms of dual asymptotic symmetries. In this sense, our result reproduces the difficulty they found. 
    
    The existence of memory that cannot be related to asymptotic symmetries is not a special property of sound. In fact, as shown by \textcite{satishchandran2019AsymptoticBehaviorMassless}, general relativity has memory effects that cannot be written in terms of a diffeomorphism. In the case of gravity, this is intimately related to how the memory tensor is decomposed in terms of scalars, vectors, and tensors on the sphere \cite{satishchandran2019AsymptoticBehaviorMassless}. Our result is astonishingly similar. Equation \eqref{eq: sound-memory-hodge-decomposed} gives the Hodge decomposition \cite{frankel2012GeometryPhysicsIntroduction,nakahara2003GeometryTopologyPhysics} of the memory tensor (see App. \ref{app: differential-forms}). It uniquely writes the memory two-form in terms of an exact piece and a harmonic piece. The exact piece is fundamentally a one-form, while the harmonic piece is fundamentally a two-form (or scalar, through Hodge duality). In the acoustic theory, ``exact memory'' is a large gauge transformation, but ``harmonic memory'' is not associated to a large gauge transformation.

    Last, but not least, we discuss the soft theorems. Within acoustics, there is once again the issue that fundamental sources are not available, and thus all sources are ``image sources.'' This introduces a difficulty in formulating what would be a feasible quantum theory in which the soft theorems should capture the memory effect. In this work, we limited ourselves to a proof of principle. If the image sources are chosen so that they have asymptotically constant monopoles, it is possible to construct a quantum field theory whose soft theorems would reproduce the memory effect. This should not be seen as a deep statement about acoustics, but rather as an application of the fact that scattering amplitudes can be used as a computational technique even in classical physics \cite{kosower2019AmplitudesObservablesClassical,cristofoli2022WaveformsAmplitudes,kosower2022SAGEXReviewScattering,mohanty2023GravitationalWavesQuantum}. The connection between these soft theorems and the asymptotic symmetries we identified had already been made by \textcite{campiglia2018CanScalarsHave,campiglia2019ScalarAsymptoticCharges,francia2018TwoFormAsymptoticSymmetries}.

    There are many development directions in sight. First and foremost, we comment on the approximations we did when describing fluid flow. Our calculations mostly concerned linear perturbations in inviscid, irrotational, and barotropic fluid flows. It would be interesting to improve on all of these approximations. The key points in which the irrotational and barotropic assumptions enter the calculations are in obtaining the equations of motion for the three-form \(\mathcal{H}\), in Sec. \ref{sec: sound-as-gauge}. To obtain the expression \(\dd\hodge\mathcal{H} = 0\), we chose to ignore these effects. Further details and more general flows can be incorporated by modifying the equations of motion to allow sources for \(\mathcal{H}\)---see, for example, Ref. \cite{fischer1999MotionQuantizedVortices}. Handling nonlinearities can lead to several modifications on the framework \cite{datta2022AnalogueGravitationalField,pal2024QuantumNonlinearEffects}, but the prediction by \textcite{datta2022InherentNonlinearityFluid} of a nonlinear acoustic memory effect suggests the infrared sector would not be greatly affected---in particular, it seems natural to speculate that asymptotic symmetries would still be present, as they are in gravity.

    Finally, in this work we focused on a linear memory effect, and considered a simple ``toy soft theorem'' as a proof of principle that an infrared triangle is available in linear acoustics. Natural development directions would be to generalize our analysis to consider the nonlinear memory effect predicted by \textcite{datta2022InherentNonlinearityFluid}, and/or the soft phonon theorems derived by \textcite{cheung2023SoftPhononTheorems}.

    Part of the calculations in this work were carried out with the aid of \textsc{Mathematica 15.0} \cite{wolframresearch2026Mathematica150}, and in particular the \textsc{OGRe} package \cite{shoshany2021OGReObjectOrientedGeneral} and the \textsc{xAct} suite \cite{martin-garciaXActEfficientTensor,martin-garcia2008XPermFastIndex}. Ref. \cite{stein2018CalculateCCESystem} was helpful in rewriting the various wave equations in Bondi coordinates. The main notebooks are available as supplemental material accompanying this preprint. 

\begin{acknowledgments}
    The work of NAA was supported by the São Paulo Research Foundation (FAPESP) under grant 2025/05161-0.
\end{acknowledgments}

\appendix
\section{Differential Forms}\label{app: differential-forms}
    For convenience, we summarize in this appendix the conventions and results we use when handling differential forms. Further details can be found in Refs. \cite{frankel2012GeometryPhysicsIntroduction,gorodski2020SmoothManifolds,lechner2018ClassicalElectrodynamicsModern,lee2012IntroductionSmoothManifolds,nakahara2003GeometryTopologyPhysics,tu2011IntroductionManifolds,wald1984GeneralRelativity}, for example. 

    Let \(M\) be a smooth, \(n\)-dimensional manifold. A differential \(p\)-form (or simply a \(p\)-form) is a smooth tensor field \(\tensor{\omega}{_{a_1}_{\cdots}_{a_p}}\) of type \((0,p)\) that is completely antisymmetric, 
    \begin{equation}
        \tensor{\omega}{_{a_1}_{\cdots}_{a_p}} = \tensor{\omega}{_[_{a_1}_{\cdots}_{a_p}_]}.
    \end{equation}
    Above, the square brackets denote antisymmetrization according to 
    \begin{equation}
        \tensor{\omega}{_[_{a_1}_{\cdots}_{a_p}_]} = \frac{1}{p!} \sum_{\pi} \mathrm{sign}(\pi) \tensor{\omega}{_{a_{\pi(1)}}_{\cdots}_{a_{\pi(p)}}},
    \end{equation}
    where the sum runs over all permutations \(\pi\) of \(p\)-elements and \(\mathrm{sign}(\pi)\) denotes the parity of the permutation. As examples, one has
    \begin{subequations}
        \begin{gather}
            \tensor{\omega}{_[_a_b_]} = \frac{1}{2}\qty(\tensor{\omega}{_a_b} - \tensor{\omega}{_b_a}), \\
            \intertext{and}
            \tensor{\omega}{_[_a_b_c_]} = \frac{1}{6}\qty(\tensor{\omega}{_a_b_c} + \tensor{\omega}{_b_c_a} + \tensor{\omega}{_c_a_b} - \tensor{\omega}{_b_a_c} - \tensor{\omega}{_a_c_b} - \tensor{\omega}{_c_b_a}).
        \end{gather}
    \end{subequations}
    Because the index structure of a \(p\)-form is trivial, we may often omit the indices altogether. For instance, we may simply write \(\omega\) instead of \(\tensor{\omega}{_{a_1}_{\cdots}_{a_p}}\).

     The space of \(p\)-form fields is denoted \(\Omega^p(M)\). \(\Omega^0(M)\) is understood as \(\mathcal{C}^{\infty}(M)\), the space of smooth functions on the manifold. Notice that, for an \(n\)-dimensional manifold \(M\), \(\Omega^{n+1}(M)\) is trivial, because it is impossible to write a nonvanishing antisymmetric tensor field with \(n+1\) indices. Similarly, all elements of \(\Omega^{n}(M)\) can be written as a function multiplying a standard basis element \(\tensor{\epsilon}{_{a_1}_\ldots_{a_n}}\).

    There are some natural operations that can be defined among \(p\)-forms. For example, take the tensor product of two forms. The result may not be antisymmetric, so we antisymmetrize. Up to a multiplicative constant, the outcome is known as the wedge product, 
    \begin{equation}
        \tensor{(\omega \wedge \mu)}{_{a_1}_{\cdots}_{a_{p}}_{b_1}_{\cdots}_{b_q}} = \frac{(p+q)!}{p!q!}\tensor{\omega}{_[_{a_1}_{\cdots}_{a_{p}}} \tensor{\mu}{_{b_1}_{\cdots}_{b_q}_]}.
    \end{equation}
    The numerical coefficient is a matter of convention.

    Even in the absence of a metric or any other extra structure, it is possible to introduce a derivative operator among differential forms. This is known as the exterior derivative \(\dd\). For each \(p\), we define \(\dd \colon \Omega^p(M) \to \Omega^{p+1}(M)\) by 
    \begin{equation}
        \tensor{\qty(\dd{\omega})}{_{a_1}_\cdots_{a_{p+1}}} = (p+1) \tensor{\nabla}{_[_{a_1}}\tensor{\omega}{_{a_2}_\cdots_{a_{p+1}}_]},
    \end{equation}
    where \(\tensor{\nabla}{_a}\) is any (torsionless) differential operator on the manifold. The antisymmetrization cancels the effect of any connection coefficients, and thus the choice of differential operator does not affect the result. In particular, one can merely use partial derivatives.

    Notice that the antisymmetrization in the definition of the exterior derivative ensures it is nilpotent. More specifically, notice that 
    \begin{subequations}
        \begin{align}
            \tensor{(\dd[2]{\omega})}{_{a_1}_\cdots_{a_{p+2}}} &= (p+2) \tensor{\partial}{_[_{a_1}}\tensor{(\dd{\omega})}{_{a_2}_\cdots_{a_{p+2}}_]}, \\
            &= (p+2)(p+1) \tensor{\partial}{_[_{a_1}}\tensor{\partial}{_{a_2}}\tensor{\omega}{_{a_3}_\cdots_{a_{p+2}}_]}, \\
            &= 0,
        \end{align}
    \end{subequations}
    because partial derivatives in any particular coordinate system always commute. Hence \(\dd[2] = 0\).

    \subsection{De Rham cohomology}
        Since \(\dd[2] = 0\), it is true that 
        \begin{equation}
            \dd\dd\omega = 0
        \end{equation}
        for any \(p\)-form \(\omega\). Suppose, however, that \(\mu\) is a \((p+1)\)-form with
        \begin{equation}
            \dd\mu = 0.
        \end{equation}
        How can we know if there is a \(p\)-form \(\omega\) such that \(\mu = \dd\omega\)? 

        This problem is addressed by de Rham cohomology---see, e.g., Refs. \cite{frankel2012GeometryPhysicsIntroduction,gorodski2020SmoothManifolds,lee2012IntroductionSmoothManifolds,nakahara2003GeometryTopologyPhysics,tu2011IntroductionManifolds} for various introductions. Let us first notice that the exterior derivative allows us to write a sequence of maps according to 
        \begin{equation}
            0 \rightarrow \Omega^0(M) \stackrel{\dd_0}{\rightarrow} \Omega^1(M) \stackrel{\dd_1}{\rightarrow} \cdots \stackrel{\dd_{n-1}}{\rightarrow} \Omega^{n}(M) \rightarrow 0,
        \end{equation}
        where we wrote \(\dd_p \colon \Omega^{p}(M) \to \Omega^{p+1}(M)\) for clarity of which map is which. Because all \((n+1)\)-forms in an \(n\)-dimensional manifold vanish, the exterior derivative applied to \(\Omega^{n}(M)\) maps all \(n\)-forms to the zero \((n+1)\)-form.

        Two spaces are particularly interesting here. One of them is the kernel 
        \begin{equation}
            Z^p(M) = \mathrm{Ker}\,\dd_p = \qty{\omega \in \Omega^p(M) \mid \dd\omega = 0}.
        \end{equation}
        The second is the range
        \begin{equation}
            B^p(M) = \mathrm{Ran}\,\dd_{p-1} = \qty{\dd\omega \mid \omega \in \Omega^{p-1}(M)}.
        \end{equation}

        Notice that \(B^p(M) \subseteq Z^p(M) \subseteq \Omega^p(M)\). The elements of \(Z^p(M)\) are said to be closed. The elements of \(B^p(M)\) are said to be exact. We want to understand when a closed form fails to be exact. This means we want to understand what are the elements of the quotient 
        \begin{equation}
            H^p(M) = Z^p(M)/B^p(M),
        \end{equation}
        which is known as the \(p\)-th de Rham cohomology group for the manifold \(M\). Although they are called groups, they are real vector spaces. 

        Notice that we defined the de Rham cohomology groups using the differentiable structure of the manifold. Indeed, we needed to define smooth differential forms in order to construct them. In spite of this, the so-called de Rham theorem ensures the de Rham cohomology groups are topological invariants \cite{frankel2012GeometryPhysicsIntroduction,nakahara2003GeometryTopologyPhysics}. 

        Let us take Euclidean space as an example. The de Rham cohomology groups are \cite{frankel2012GeometryPhysicsIntroduction,nakahara2003GeometryTopologyPhysics,tu2011IntroductionManifolds}
        \begin{equation}
            H^k(\mathbb{R}^n) = \begin{cases}
                \mathbb{R}, & \text{ if } k = 0, \\
                0, & \text{ if } k > 0,
            \end{cases}
        \end{equation}
        where ``\(0\)'' is used in the second line to denote the trivial vector space. Hence, in Euclidean space, all closed forms are exact. This is known as Poincaré's lemma. On a manifold, one can always pick a local coordinate chart with the topology of Euclidean space. It then follows that, in this chart, all closed forms are exact. Hence, in a general manifold, all closed forms are locally exact, meaning it is always possible to write a closed form \(\mu\) as \(\mu = \dd\omega\) on a sufficiently small region.

        Our main case of interest is that of a two-sphere. One then has that \cite{frankel2012GeometryPhysicsIntroduction,nakahara2003GeometryTopologyPhysics}
        \begin{subequations}
            \begin{align}
                H^0(\mathbb{S}^2) &= \mathbb{R}, \\
                H^1(\mathbb{S}^2) &= 0, \\
                H^2(\mathbb{S}^2) &= \mathbb{R}.
            \end{align}
        \end{subequations}
        In fact, \(H^0(M) = \mathbb{R}\) whenever \(M\) is connected, because the constant \(0\)-forms are closed, but cannot be exact, since \(B^0(M) = 0\)---there are no \((-1)\)-forms. The remaining groups are specific to the sphere. Notice 
        \begin{equation}
            \mathrm{dim}\,H^2(\mathbb{S}^2) = 1,
        \end{equation}
        where the dimension is meant in the sense of real vector spaces. In this sense, there is a single two-form that is closed, but not exact. This is the volume two-form on the sphere, 
        \begin{equation}
            \epsilon = \sin\theta \dd{\theta} \wedge \dd{\varphi},
        \end{equation}
        where \(\theta\) is the polar angle and \(\varphi\) is the azimuthal angle. Notice we say that ``there is a single two-form that is closed, but not exact'' in the sense that, if \(\omega \in \Omega^2(\mathbb{S}^2)\) is closed, but not exact, then it always holds that
        \begin{equation}
            \omega = \dd\mu + c \epsilon,
        \end{equation}
        where \(\mu \in \Omega^1(\mathbb{S}^2)\) and \(c\) is a real number.

        In the main text, we often work in the ``radiation zone,'' modeled as the manifold \(\mathbb{R}^2 \times \mathbb{S}^2\). It can be shown that 
        \begin{equation}
            H^p(\mathbb{R}^2 \times \mathbb{S}^2) = H^p(\mathbb{S}^2)
        \end{equation}
        for all \(p\) (this follows from the Poincaré lemma and the so-called Künneth formula \cite{frankel2012GeometryPhysicsIntroduction}).

    \subsection{Hodge theory}
        As mentioned above, all \(n\)-forms in an \(n\)-dimensional manifold can be written as a function multiplying a ``standard form'' \(\epsilon\). If we are working in a pseudo-Riemannian manifold, then the choice of a ``standard form'' is specified up to a choice of sign. More specifically, it is convenional to pick
        \begin{equation}
            \tensor{\epsilon}{^{a_1}^{\cdots}^{a_n}}\tensor{\epsilon}{_{a_1}_{\cdots}_{a_n}} = (-1)^s n!,
        \end{equation}
        where indices are raised with the metric and \(s\) is the number of negative signs in the metric signature. It then follows that \cite{wald1984GeneralRelativity}
        \begin{equation}
            \tensor{\epsilon}{^{a_1}^{\cdots}^{a_n}}\tensor{\epsilon}{_{b_1}_{\cdots}_{b_n}} = (-1)^s n! \tensor{\delta}{^[^{a_{1}}_{b_{1}}} \cdots \tensor{\delta}{^{a_n}^]_{b_n}}.
        \end{equation}
        If we contract some of the indices, we obtain
        \begin{multline}
            \tensor{\epsilon}{^{a_1}^{\cdots}^{a_i}^{a_{i+1}}^{\cdots}^{a_n}}\tensor{\epsilon}{_{a_1}_{\cdots}_{a_i}_{b_{i+1}}_{\cdots}_{b_n}} \\ = (-1)^s (n-i)! i! \tensor{\delta}{^[^{a_{i+1}}_{b_{i+1}}} \cdots \tensor{\delta}{^{a_n}^]_{b_n}}.
        \end{multline}
        We note that the existence of a global choice of sign depends on topological properties of the manifold, but this is not relevant for our purposes.

        Once a choice of sign for \(\epsilon\) has been made, it is possible to define an operator \(\hodge \colon \Omega^{p}(M) \to \Omega^{n-p}(M)\) known as the Hodge dual. We define
        \begin{equation}
            \tensor{(\hodge \omega)}{_{a_1}_{\cdots}_{a_{n-p}}} = \frac{1}{p!} \tensor{\omega}{^{b_1}^{\cdots}^{b_p}}\tensor{\epsilon}{_{b_1}_{\cdots}_{b_p}_{a_1}_{\cdots}_{a_{n-p}}},
        \end{equation}
        where indices are raised with the metric. Notice that
        \begin{equation}
            \hodge\hodge\omega = (-1)^{s+p(n-p)} \omega.
        \end{equation}

        Using the Hodge dual and the exterior derivative, we can define the codifferential \(\var \colon \Omega^{p}(M) \to \Omega^{p-1}(M)\) through
        \begin{equation}
            \var = (-1)^{s+1+n(p+1)} \hodge \dd \hodge = (-1)^{p} \hodge^{-1} \dd \hodge.
        \end{equation}
        Notice \(\hodge^{-1}\) is acting on a \((n-p+1)\)-form. For Lorentzian manifolds in four dimensions we get
        \begin{equation}
            \var = \hodge \dd \hodge.
        \end{equation}
        For Riemannian manifolds in two dimensions we get
        \begin{equation}
            \var = - \hodge \dd \hodge.
        \end{equation}
        In the main text, we prefer to write \(\dd\) and \(\hodge\) explicitly and \emph{never} write \(\delta\) for the codifferential. This avoids mistaking it with the variation or perturbation of various quantities. Nevertheless, understanding the codifferential can simplify some expressions. As a general rule, it holds that
        \begin{equation}
            \tensor{\qty(\var{\omega})}{_{a_1}_\cdots_{a_{p-1}}} = - \tensor{\nabla}{^b}\tensor{\omega}{_{b}_{a_1}_\cdots_{a_{p-1}}},
        \end{equation}
        and thus it is akin to the divergence of a differential form, while the exterior derivative is similar to the curl.

        A form \(\mu\) with \(\dd\mu = 0\) is closed, and a form \(\mu\) with \(\mu = \dd\omega\) is exact. Analogously, a form \(\mu\) with \(\var\mu = 0\) is coclosed, and a form \(\mu\) with \(\mu = \var\omega\) is coexact. \(\mu\) being coclosed is equivalent to \(\hodge\mu\) being closed, and \(\mu\) being coexact is equivalent to \(\hodge\mu\) being exact.

        The codifferential and the exterior derivative can be used together to define the Laplacian on forms. This is known as the Laplace--de Rham operator, or the Hodge Laplacian. It is given by 
        \begin{equation}
            \Delta_{\text{H}} \omega = (\dd \var + \var\dd)\omega.
        \end{equation}
        This operator is related to the connection Laplacian \(\tensor{\nabla}{_a}\tensor{\nabla}{^a}\), but they are not the same. They are related by the Weitzenböck formula \cite{sanders2014ElectromagnetismLocalCovariance,derham1984DifferentiableManifoldsForms}
        \begin{multline}
            \Delta_{\text{H}}\tensor{\omega}{_{a_1}_{\cdots}_{a_p}} = - \tensor{\nabla}{^b}\tensor{\nabla}{_b}\tensor{\omega}{_{a_1}_{\cdots}_{a_p}} + p \tensor{R}{_[_{a_1}^b}\tensor{\omega}{_|_b_|_{a_2}_{\cdots}_{a_p}_]} \\ - \binom{p}{2} \tensor{R}{_[_{a_1}_{a_2}^b^c}\tensor{\omega}{_|_b_c_|_{a_3}_{\cdots}_{a_p}_]}.
        \end{multline}

        As a key example, let us consider four-dimensional electrodynamics in Lorentzian signature. The equations of motion are
        \begin{subequations}
            \begin{gather}
                \dd\mathcal{F} = 0, \\
                \dd\hodge\mathcal{F} = \hodge{}j.
            \end{gather}
        \end{subequations}
        In terms of the codifferential we have
        \begin{subequations}
            \begin{gather}
                \dd\mathcal{F} = 0, \\
                \var\mathcal{F} = j.
            \end{gather}
        \end{subequations}
        Because \(\mathcal{F}\) is closed, we can introduce a potential \(\mathcal{A}\) at least locally. This means the equations of motion become
        \begin{subequations}
            \begin{gather}
                \mathcal{F} = \dd\mathcal{A}, \\
                \var\dd\mathcal{A} = j.
            \end{gather}
        \end{subequations}
        This holds for a general gauge. In Lorenz gauge, we impose \(\dd\hodge \mathcal{A} = 0\). In terms of the codifferential, \(\var\mathcal{A} = 0\). Hence, Lorenz gauge means imposing the potential is coclosed. With this in mind, we see we can now write Maxwell's equations as
        \begin{subequations}
            \begin{align}
                \var\dd\mathcal{A} &= j, \\
                (\var\dd + \dd\var)\mathcal{A} &= j, \\
                \Delta_{\text{H}}\mathcal{A} &= j.
            \end{align}
        \end{subequations}
        By similar arguments, we see \(\var\dd\) (or \(\dd\hodge\dd\), as we write in the main text) is always understood as a wave operator when acting on coclosed forms (i.e., on fields in Lorenz gauge). Notice too that scalars are always coclosed. 
        
        For us, the case of a two-sphere is particularly interesting. Then the Riemann tensor is considerably simplified, and we find
        \begin{subequations}
            \begin{gather}
                \Delta_{\text{H}} C = - \mathcal{D}^2 C, \\
                \Delta_{\text{H}}\tensor{C}{_A} = - (\mathcal{D}^2 - 1) \tensor{C}{_A}, \\
                \Delta_{\text{H}}\tensor{C}{_A_B} = - \mathcal{D}^2\tensor{C}{_A_B},
            \end{gather}
        \end{subequations}
        where \(\mathcal{D}^2 = \tensor{\mathcal{D}}{^A}\tensor{\mathcal{D}}{_A}\) is the connection Laplacian on the sphere (as in the main text). 

        In compact Riemannian manifolds (such as the sphere), a form \(\omega\) is said to be harmonic when \(\Delta_{\text{H}}\omega = 0\). This occurs if, and only if, \(\omega\) is closed and coclosed \cite{nakahara2003GeometryTopologyPhysics,frankel2012GeometryPhysicsIntroduction}. The Hodge decomposition theorem then ensures that---in compact Riemannian manifolds---one can write
        \begin{equation}
            \Omega^{p}(M) = \dd\Omega^{p-1}(M) \oplus \var\Omega^{p+1}(M) \oplus \mathrm{Harm}^{p}(M),
        \end{equation}
        where \(\mathrm{Harm}^{p}(M)\) is the space of harmonic \(p\)-forms. Hence, all \(p\)-forms on a compact Riemannian manifold can be uniquely decomposed into a sum of an exact form, a coexact form, and a harmonic form. 

        Finally, Hodge's theorem states that---in a compact Riemannian manifold---the space of harmonic \(p\)-forms is isomorphic to the de Rham cohomology \(H^p(M)\) \cite{nakahara2003GeometryTopologyPhysics,frankel2012GeometryPhysicsIntroduction},
        \begin{equation}
            H^p(M) \cong \mathrm{Harm}^p(M).
        \end{equation}
        In this sense, the harmonic forms are precisely the forms that are closed, but fail to be exact.

\bibliography{bib}
\end{document}